\documentstyle[epsfig]{mn}
\DeclareMathVersion{bold}
\title[Filter comparison]{Comparing filters for the detection of point sources}
\author[Barreiro et al.]{R.B. Barreiro, J.L. Sanz,
D. Herranz\footnotemark  ~and 
E. Mart\'\i nez-Gonz\'alez   \\
Instituto de F\'\i sica de Cantabria, Santander, Spain\\}
\date{Accepted ???. Received ???; in original form ???}

\begin{document}
\maketitle

\begin{abstract}
This paper considers filters (the Mexican hat wavelet, the matched and
the scale-adaptive   
filters) that optimize the detection/separation of point sources on a
background. We make a one-dimensional treatment, we assume that the
sources have a Gaussian profile, i. e.  
$\tau (x) = e^{- x^2/2R^2}$, and a background modelled by
an homogeneous  
and isotropic Gaussian random field, characterised by a power spectrum
$P(q)\propto q^{-\gamma},  
\gamma \geq 0$. Local peak detection is used after
filtering. Then,  
the Neyman-Pearson criterion is used to define the confidence level
for detections and  
a comparison of filters is done based on the number of spurious and
true detections. We have performed numerical simulations to test
theoretical ideas and conclude that the results of the simulations
agree with the analytical results.
\end{abstract}


\section{Introduction}
\footnotetext{Currently at ISTI/CNR, Pisa, Italy} 
The detection of localized signals or features on one-dimensional
(1D) or two-dimensional  
images (2D) is one of the most challenging aspects of image analysis. 

We are interested in the detection of compact sources (signal)
embedded in a background 
(1D case). We assume that the profile of the source and the
statistical properties of  
the background are known. Linear filtering of the data in order to eliminate
partially the background is the 
primary goal. Several filters have been introduced in the literature
to deal with the problem. Four examples are:  
the continuous `Mexican hat' wavelet, difference of two Gaussians,
matched filters and  
scale-adaptive filters. The two first cases are filters given `a
priori', adapted to the detection  
of point sources, whereas the matched filter is constructed taking
into account the profile and 
background in order to get the maximum SNR at the source position. The
scale-adaptive filter  
is constructed taking into account the previous properties and also
the constrain to have a maximum 
in filtered space at the scale and source position. Hereinafter, we
will identify the position of the possible sources with the local maxima.

   The `Mexican hat' wavelet used as a filter has been extensively
applied during the last 
years to analyse optical, X-ray and microwave data. Optical images of
galaxy fields have been  
analysed to detect voids and high-density structures 
in the first CfA redshift survey slice (Slezak et al. 1993). Microwave 
images have been analysed (Cay\'on et al. 2000; Vielva et al. 2001a)
and combined with the maximum 
entropy method (Vielva et al. 2001b) to obtain catalogues of point
sources from simulated maps 
at different frequencies that will be observed by the future Planck
mission. On the other hand,  
the Mexican hat has also been used to detect X-ray sources (Damiani et
al. 1997) and presently for 
the on-going XMM-Newton mission (Valtchanov et al. 2001).

   Other useful filters include the so-called `matched' filters. They
optimize the 
signal-to-noise ratio and have been used mainly in signal processing. The 
generalization to two dimensions of the previous matched filters 
have been used recently to detect clusters of galaxies from optical 
imaging data (Postman et al. 1996; Kawasaki et al. 1998). In this
approach the method uses 
galaxy positions, magnitudes (and photometric/spectroscopic redshifts
if available) to find 
clusters and determine their redshift. Also they have been applied to
microwave maps to detect 
clusters through the Sunyaev-Zeldovich effect, either on single or
multifrequency maps (Herranz et al. 2002a, b)

   Other remarkable filters that have been recently (Sanz et al. 2001)
introduced in the  
astrophysics literature are the so-called scale-adaptive
filters. Applications have been done  
to the detection of point sources in time-ordered data in the future
Planck mission  
(Herranz et al. 2002c), detection of
SZ-emission in single microwave maps and X-ray emission from clusters
of galaxies in single  
X-ray maps (Herranz et al. 2002a). More recently, application for the
detection of SZ-clusters in  
multifrequency maps representing the future observations by the Planck
mission has been considered (Herranz et al. 2002b).

   One interesting question relative to all of these filters is the
$\it optimality$, that we define  
in terms of the following properties: confidence level of the
detections, number of spurious sources  
which emerge in the process and number of real sources detected
(detection limit and completitude  
magnitude). As we will see in this paper, the previous properties are
not only related to the SNR 
gained in the filtering process but depend on the filtered momenta to
4th-order, the curvature of the 
source and the amplification in the 1D case. The combination of these
quantities in a complicated way 
makes the decision on one filter totally dependent on the source
profile and the background. This shows that, in addition to the
amplification, one must take into account other quantities which also
play an important role. The amplification was
suggested by Vio et al. (2002) in order to compare filters. The final
identification of the sources  
with a simple $n\sigma $ thresholding is not the whole story
regarding optimality. Moreover, if the scale of the source needs to be  
estimated from the data, it must be identified `a posteriori'
on the SNR map when using matched filters, what introduces noise in
the process. On the contrary, the ${\it adaptive}$ filter  
allows one to get the sources straightforwardly on the filtered map.

In section 2, we introduce two useful quantities: number of maxima
in a Gaussian background and  
define the detection problem. In section 3, we remark on the optimal
statistic that allows one to define the 
region of acceptance, i. e. the confidence level of the detections. In
section 4, we comment on the 
different filters to be compared (Mexican hat, matched and
scale-adaptive filters). In section 5, we 
give some analytical and numerical results. In section 6, we
describe the numerical simulations performed to test some theoretical
aspects and give the main 
results. Finally, in section 7, we summarize the main results and
applications of this paper. 

\section{Local peak detection}

Let us assume a 1D background (e. g. one-dimensional scan on the
celestial sphere or time 
ordered data set) represented by a Gaussian random field $\xi (x)$
with average value  
$\langle \xi (x)\rangle = 0$ and power spectrum $P(q), q\equiv |Q|$: 
$\langle \xi (Q)\xi^* (Q')\rangle = P(q)\delta_D (q - q')$,
$\xi (Q)$ is the Fourier transform of $\xi (x)$ and $\delta_D$ is the 1D Dirac 
distribution. 
The distribution of
maxima was studied by Rice (1954) in a pioneer article, the expected
number density of 
maxima per intervals $(x, x + dx)$, $(\nu ,\nu + d\nu )$ and $(\kappa
,\kappa + d\kappa )$ is given by
\begin{equation}
n_b(\nu ,\kappa )  = \frac{n_b}{\sqrt{2\pi} }\frac{\kappa}{\sqrt{1-\rho^2}} 
e^{- \frac{1}{2(1 - \rho^2)}(\nu^2 + \kappa^2 - 2\rho \nu \kappa)}, 
\label{nbackground}
\end{equation}
\begin{displaymath}
n_b \equiv \frac{1}{2\pi \beta \sqrt{\rho}},~
\nu \equiv \frac{\xi}{\sigma_0},~ \kappa \equiv \frac{-\xi^{\prime
\prime}}{\sigma_2},~ 
\beta \equiv  \sqrt{\frac{\sigma_0}{\sigma_2}},~
 \rho \equiv \frac {\sigma_1^2}{\sigma_0 \sigma_2},~ \nonumber \\
\end{displaymath}
where $\nu \in (-\infty ,\infty )$ and $\kappa \in (0,\infty )$ represent the 
normalized field and curvature, 
respectively, and $n_b$ is the expected total number density of maxima
(i. e. number of 
maxima per unit interval $dx$). $\sigma_n^2$ is the moment of order
$2n$ associated to the field. 

If the original field is linear-filtered with a circularly-symmetric 
filter $\Psi (x; R, b)$, dependent on
$2$ parameters ($R$ defines a scaling whereas $b$ defines a translation)
\begin{equation}
\Psi (x; R, b) = \frac{1}{R}\psi \left(\frac{|x - b|}{R}\right),
\end{equation}
we define the filtered field as
\begin{equation}
w(R, b) = \int dx\,\xi (x)\Psi (x; R, b).
\end{equation}
Then, the moment of order $n$ of the linearly-filtered field is
\begin{equation}
\sigma_n^2 \equiv 2\int_0^{\infty} dq\,q^{2n}P(q)\psi^2 (Rq),
\end{equation}
being $P(q)$ the power spectrum of the unfiltered field and $\psi (Rq)$ 
the Fourier transform of the circularly-symmetric linear filter.

Now, let us consider a Gaussian source (i. e. profile given by 
$\tau (x) = e^{- x^2/2R^2}$) embedded in the previous
background. Then,  
the expected number density of maxima per intervals $(x, x + dx)$,
$(\nu ,\nu + d\nu )$ and  
$(\kappa ,\kappa + d\kappa )$, given a source of amplitude $A$ in such
spatial interval,  is given by
\begin{eqnarray}
n(\nu ,\kappa |\nu_s) &=& \frac{n_b}{\sqrt{2\pi}}\frac{\kappa}{\sqrt{1 -
\rho^2}} \nonumber \\ 
&& e^{- \frac{(\nu - \nu_s)^2 + (\kappa -
\kappa_s)^2 -  2\rho (\nu - \nu_s)(\kappa - \kappa_s)}{2(1 - \rho^2)}},
\label{nsource}
\end{eqnarray}
where $\nu \in (-\infty ,\infty )$ and $\kappa \in (0,\infty )$, 
$\nu_s = A/\sigma_0$ is the normalized amplitude of the source and 
$\kappa_s = - A\tau_{\psi}^{\prime \prime}/\sigma_2$ is the
normalized curvature of the filtered source.
The last expression can be obtained as
\begin{equation}
\kappa_s = \nu_s y_s,~ y_s \equiv - \beta^2 \tau_{\psi}^{\prime
\prime},~
- \tau_{\psi}^{\prime \prime} = 2\int_0^{\infty}dq\,q^2\tau (q)\psi (Rq).
\end{equation}
We consider that the filter is normalized such that the amplitude of
the source is the same after linear filtering: $\int dx\,\tau (x)\Psi
(x; R, b) = 1$. 

The total number density $n$ in the presence of a local source is
obtained integrating eqn.~(\ref{nsource})
\begin{equation} 
n = n_be^{-\frac{\kappa_s^2}{2}}\left[1 +
B\left(\frac{\kappa_s}{\sqrt{2}}\right)\right],~~
B(x)\equiv \sqrt{\pi}xe^{x^2}erfc(- x).
\label{totalnsource}
\end{equation}

\section{Optimal statistics}
We want to make a decision between filters based on $\it
optimality$. We will assume that it 
includes the following properties: a) confidence level of the
detections, b) number of spurious  
sources which emerge in the process and c) number of real sources
detected. As we will see, the  
previous properties are not only related to the SNR gained in the
filtering process but depend on  
the filtered momenta to 4th-order, the curvature of the source and the
amplification (1D case). 

We will distinguish two cases: I) {\it Simple detection},
consisting of detecting the presence  
of a signal $s(x)$ in a background and II) {\it Simple measurement},
consisting of detecting the  
signal $s(x) = A\tau (x)$ and measuring its unknown amplitude $A$ (we
assume the profile is given)  
in the presence of a background.
  
\subsection{Simple detection} \label{sec:simple_detection}

First, let us establish the confidence level of the detections
assuming $\it simple$ detection (the 
amplitude of a source $A$ is given and so we can calculate $\nu_s$):
let us consider a local peak in  
the 1D data set characterised by the normalized amplitude and
curvature $(\nu ,\kappa)$, 
if $H_0: p.d.f.\, p(\nu ,\kappa |\nu_s )$ represents the $\it null$
hypothesis that the peak is a  
source with normalized amplitude $\nu_s$ given the data $(\nu ,\kappa)$, and 
$H_1: p.d.f.\, p(\nu ,\kappa |0)$ represents the $\it alternative$
hypothesis that the peak is a  
maximum of the background, we can associate to any region $R_*(\nu
,\kappa )$ two errors 
\begin{equation}
\alpha = \int_{R_*}d\nu \,d\kappa \,p(\nu ,\kappa |0),~~
1 - \beta = \int_{R_*}d\nu \,d\kappa \,p(\nu ,\kappa |\nu_s ),
\label{alfaybeta}
\end{equation}
$\alpha$ is the false alarm probability or confidence level
of the detection (i. e.  it
represents the probability of interpreting noise as signal) whereas
$\beta$ is the false dismissal 
probability or equivalently, $1 - \beta$ is the power of the detection
(i. e. $\beta$ represents the  
probability of interpreting signal as noise). $R_*$ is called the
$\it acceptance$ region. 
 
Clearly, the previous probabilities can be obtained from the number
density of maxima given by 
eqns. (\ref{nbackground},~\ref{nsource},~\ref{totalnsource}) as  
\begin{equation}
p_b(\nu ,\kappa )\equiv  p(\nu ,\kappa |0) = \frac{n_b(\nu ,\kappa
)}{n_b},~~
p(\nu ,\kappa |\nu_s ) =  \frac{n(\nu ,\kappa |\nu_s)}{n}, 
\label{pbyp}
\end{equation}
where $n$ and $n_b$ are the total number density in the
presence or absence of a local source, respectively.
   
   We will assume as decision rule the Neyman-Pearson one (C1): the
acceptance region $R_*$ giving the 
highest power $1 - \beta$, for a given confidence level $\alpha$, is the region
\begin{equation}
L(\nu ,\kappa |\nu_s)\equiv \frac{p(\nu ,\kappa |\nu_s )}{p(\nu
,\kappa |0)}\geq L_*, 
\label{lik_nus}
\end{equation}
where $L_*$ is a constant. So, the decision rule is
expressed by the likelihood ratio: if 
$L\geq L_*$ the signal is present, whereas if $L< L_*$ the signal is absent.
Once we have assumed the previous decision rule for simple detection,
one can calculate the ROC-curves  
(`receiver operating curves' in the signal processing jargon): $\beta
= \beta (\alpha ;\nu_s )$. 

Now, we will introduce the significance $s(\nu_s)$ of the detection
\begin{equation}
s^2\equiv \frac{{[\langle N\rangle_{signal} -  \langle
N\rangle_{no-signal}]}^2} 
{\sigma^2_{signal} + \sigma^2_{no-signal}},
\label{eq_significance}
\end{equation}
where in the numerator appears the difference between the
mean number of peaks in the  
presence and absence of signal: $(1 - \beta )N$ and $\alpha N$,
respectively. In the 
denominator appear the variances in the presence and absence of
signal: $\beta (1 - \beta )N$ and 
$\alpha (1 - \alpha )N$, respectively. The last quantities have been
calculated taken into account that,  
in the absence of signal, the probability of detecting locally peaks
in $q$ of $N$  
realisations of the background is given by the binomial distribution: 
$P_q = \left(\begin{array}{cc} N \\ q \\ \end{array} \right)\alpha^q
(1 - \alpha )^{N - q}$. Therefore, the 
significance is given by the function 
\begin{equation}
s^2(\alpha )\propto \frac{{(1 - \beta - \alpha )}^2}{\beta (1 -  \beta
) + \alpha (1 -  \alpha )}, 
\end{equation}
The next step is to assume a $\it criterion$ to define
the optimal confidence level (C2): maximizing $s$ respect to $\alpha$
(or equivalently $L_*$). 
In this sense, for each concrete filter we are able to get the best
conditions (i. e.  
the maximum power corresponding to the optimal confidence level in the
sense of maximizing the 
significance of the detection, see Allen et al. 2002) for the filtered
data to be analysed. Clearly,  
for each filter we have found a unique region of acceptance $R_*$
given the amplitude of the source $A$.

\subsection{Simple measurement} \label{sec:simple_measurement}

In this case the signal has an unknown parameter, the amplitude, that
is measured. First, we are 
interested in the detection of the signal and then in the estimation
of its amplitude. Therefore, 
we calculate the a posteriori p.d.f. of all the possible values of the
parameter $A$ given the data 
$(\nu ,\kappa )$. In absence of
any a priori information about the p.d.f. $p(A)$ we will assume
uniformity in the 
interval $[0, \nu_c ]$, i. e. we can integrate over $\nu_s$: 
$p(\nu ,\kappa ) =  \frac{1}{\nu_c}\int_0^{\nu_c}d\nu_s \,p(\nu
,\kappa |\nu_s )$. 
With this new p.d.f., the likelihood ratio is defined as
\begin{equation} \label{eq:likelihood_se}
L(\nu ,\kappa )\equiv \frac{p(\nu ,\kappa )}{p(\nu ,\kappa |0)} = 
\frac{1}{\nu_c}\int_0^{\nu_c}d\nu_s \,L(\nu ,\kappa |\nu_s ),
\end{equation}
%
%where we have assumed the cut-off $\nu_c = (\xi_{max} -
%\xi_{min})/\sigma_0$, being 
%$\xi_{max}$, $\xi_{min}$ the maximum and minimum values of the field
%in the original image. 
We express the decision rule giving the region of acceptance $R_*$ as
(criterion C1) 
\begin{equation} \label{eq:acceptance_se}
R_*: L(\nu ,\kappa )\geq L_*,
\end{equation}
where $L_*$ is a constant. Once we have assumed the previous decision rule for 
{\it simple measurement}, one can calculate the ROC-curves $\beta =
\beta (\alpha )$, 
where one must use $p_b(\nu ,\kappa )$ and $p(\nu ,\kappa )$, respectively. By 
introducing the significance $s$ of the detection by
eqn.~(\ref{eq_significance}), we adopt the same C2 criterion: 
maximizing $s$ respect to $\alpha$, to obtain the optimal confidence level.

\noindent Moreover, the number of spurious (corresponding to
background fluctuations) sources $n^*_b$ 
and real detections $n^*$ are obtained by integration of $n_b$ and
$n(\nu,\kappa)\equiv \frac{1}{\nu_c}\int_{0}^{\nu_c} d\nu_s
n(\nu,\kappa | \nu_s)$ 
in the region of acceptance $R_*$.

On the other hand, regarding the estimation of the amplitude $A$, it
is natural to define the most probable value of
$A$ (i.e. the amplitude $A$ corresponding to the value ${\nu}_s$ where
the likelihood $L(\nu ,\kappa |\nu_s )$  takes its maximum)
as being the result 
of the measurement and consider this value as the $\it measured$ value of $A$. 

\section{The filters}
\subsection{The scale-adaptive filter (SAF)}

The idea of a scale-adaptive filter (or optimal pseudo-filter) has been
recently introduced by the authors (Sanz et al. 2001).
By introducing a circularly-symmetric filter, $\Psi (x; R, b)$, 
we are going to express the conditions in order to obtain a
scale-adaptive filter for the 
detection of the source $s(x)$ at the origin taking into account the
fact that the source is 
characterised by a single scale $R_o$. The following conditions are assumed: 
$(1) \langle w(R_o, 0)\rangle = s(0) \equiv A$, i. e. $w(R_o, 0)$ is an ${\it
unbiased}$ estimator of the amplitude of the source. 
$(2)$ the variance of $w(R, b)$ has a minimum at the scale $R_o$,
i. e. it is an ${\it efficient}$ estimator.
$(3)$  $w(R, b)$ has a maximum at $(R_o, 0)$.
Then, the filter satisfying these conditions is given by the equation
\begin{equation} \label{eq:saf}
\tilde{\psi} (q) \equiv \psi (R_oq) = \frac{1}{ac - b^2} \frac{\tau (q)}{P(q)}
\left[nb + c - (na + b)\frac{dln\tau}{dlnq}\right],
\end{equation}
\begin{equation} 
a\equiv \int dq\,\frac{{\tau}^2}{P},~~
b\equiv \int dq\,q \frac{\tau}{P}\frac{d\tau}{dq},~~
c\equiv \int dq\,q^2\frac{1}{P}{\left(\frac{d\tau}{dq}\right)}^2.
\end{equation}
where $\tau$ is the profile of the source in Fourier space
($s(x) = A\tau (x)$).
Generically, $\Psi$ is not positive (i. e. the name ${\it
pseudo-filter}$) and $\Psi$ 
does not define a continuous wavelet transform. Moreover, the filter
${\it adapts}$ 
to the source profile, the background and the scale of the source,
i. e. the name ${\it 
adaptive}$ filter.

The previous equations have been used by the authors to obtain the
adaptive filter for a Gaussian 
and an exponential profile (Sanz et al. 2001) and a multiquadric
profile (Herranz et al. 2002a,b). 
The previous SAF has been recently used by Herranz et al. (2002c) for
point source  
detection and extraction from simulated Planck time-ordered data. 

Assuming a scale-free power spectrum, $P(q) = Dq^{- \gamma}$, and a
Gaussian profile for the source, the previous set of equations lead to
the filter 
\begin{equation}
\psi_a (q) = \frac{1}{\Gamma (m)}q^{\gamma}e^{- \frac{1}{2}q^2}\left[1
- t +  \frac{t}{m}q^2\right],
\end{equation}
\begin{displaymath}
m\equiv \frac{1 + \gamma}{2},~~ t\equiv \frac{1 - \gamma}{2}.
\end{displaymath}
\begin{figure}
\centerline{\epsfig{file=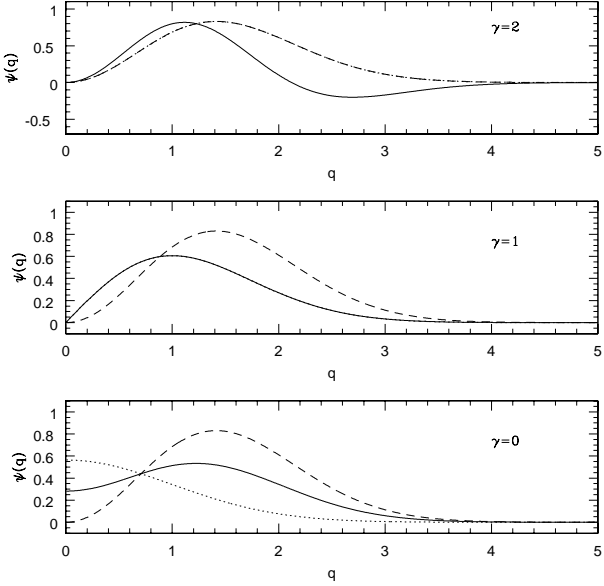, angle=0, width=\hsize}}
\caption{Scale-adaptive filter (solid line), matched filter (dotted 
line) and Mexican Hat wavelet (dashed line) in Fourier space for the
cases $\gamma=0$, $\gamma=1$ and $\gamma=2$. The scale parameter $R$
for the filters is taken to be $R=1$. Note that for $\gamma=1$ the
adaptive and matched filter coincide, whereas the matched filter and
the Mexican Hat wavelet are equal for $\gamma=2$.}
\label{filters}
\end{figure}
In Fig.\ref{filters} appears the SAF for different values of
the spectral index $\gamma = 0, 1, 2$.
In this case the filter parameters $\beta$ and $\rho$ and the
curvature of the source $y_s$ are given by
\begin{eqnarray}
\beta^2 = R^2 \sqrt{\frac{1 +\frac{t^2}{m}}{m(1+m)\left[1 + \frac{t^2}{m} +  
\frac{2t(2 + t)}{m^2}\right]} },~~~~ 
y_s = \frac{\beta^2}{R^2}, \nonumber \\
\rho = \sqrt{\frac{m}{1+m}} 
\frac{1 + \frac{t^2}{m} + \frac{2t}{m^2}}{\sqrt{
\left(1 + \frac{t^2}{m}\right)\left(1 + \frac{t^2}{m} + \frac{2t(2 +
t)}{m^2}\right)}}
\end{eqnarray}
\begin{figure}
\centerline{\epsfig{file=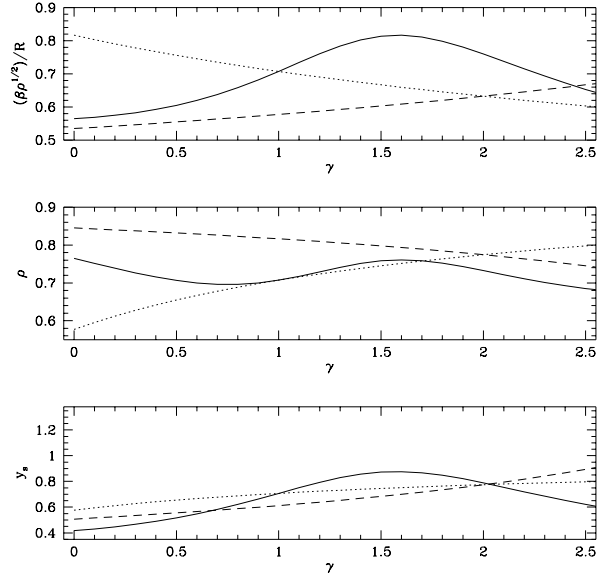, angle=0,width=\hsize}}
\caption{$\beta\sqrt{\rho}, \rho$ and $y_s$ as a function of the spectral 
index $\gamma$ for the SAF (solid line), MF (dotted line) and MHW
(dashed line).}
\label{parameters}
\end{figure}
In Fig.~\ref{parameters} appear the parameters
$\beta\sqrt{\rho}$, $\rho$ and $y_s$ as a function of the  
spectral index $\gamma $ for the SAF.

The identification of sources as peaks above a certain threshold
(e. g. $3\sigma_0$) in filter space gives a low probability of false
detections (reliability) because if the 
background has a characteristic scale of variation different from the
sources, then practically everything detected with our method is real. 

For comparison with the filter developed in this subsection, we shall
briefly introduce 
other couple of filters that have been extensively used in the
literature: the Mexican hat
wavelet and the matched filter.

\subsection{The Mexican Hat wavelet (MHW)}

The MHW on $\Re$ is defined to be proportional to the Laplacian of the
Gaussian function: 
$\psi_h (x) \propto (1 - x^2) e^{- x^2/2}$. Thus, in Fourier space
\begin{equation}              
\psi_h (q) = \frac{2}{\pi^{1/2}} q^2e^{- \frac{1}{2}q^2}.
\end{equation}
In Fig.~\ref{filters} appears the MHW compared to other filters.

In this case the filter parameters $\beta$ and $\rho$ and the
curvature of the source $y_s$ are given by
\begin{equation}
\beta^2 = \frac{R^2}{\sqrt{(2+t)(3+t)}},~~\rho = \sqrt{\frac{2+t}{3+t}},~~
y_s = \frac{3\beta^2}{2R^2}.
\end{equation}
\noindent The parameters $\beta\sqrt{\rho} ,\rho$
and $y_s$ as a function of the spectral index $\gamma$ 
are given in Fig.~\ref{parameters} for the MHW.

\noindent We comment that the generalization of this type of wavelet
for two dimensions  
has been extensively used for point source detection in 2D images.

It should be noted that the MHW studied in this work is constructed at 
a fixed scale $R$ given by the profile of the source. This differs
from the filter used in Vielva et al (2001a,b) where an optimal scale
is determined from the data and used to construct the MHW. This
optimal scale is chosen to give maximum amplification of the source
and thus the MHW at the optimal scale (MHW$_o$) will give a higher gain 
than the MHW at the scale of the source. Moreover, Vielva et al. perform a
multiscale fit in order to estimate the amplitude of the source.

\subsection{The matched Filter (MF)}

If one removes condition (3) defining the SAF
in the previous section, it is not difficult to find another type of
filter after minimization of the variance (condition  
(2)) with the constrain (1)
\begin{equation} \label{eq:mf}
\tilde{\psi}_m (q) \equiv \psi (R_oq) = \frac{1}{2a}\frac{\tau (q)}{P(q)}.
\end{equation} 
\noindent This will be called ${\it matched}$ filter as is usual in
the literature. Note that in general the matched and adaptive filters
are different.

For the case of a Gaussian profile for the source and a scale-free power 
spectrum given by $P(q)\propto q^{-\gamma}$, the previous formula leads to the 
following matched filter
\begin{equation} 
\psi_m (q) = \frac{1}{\Gamma (m)}q^{\gamma}e^{- \frac{1}{2}q^2},\ \ \ 
m\equiv \frac{1 + \gamma}{2}.
\end{equation}
In Fig.~\ref{filters} appears the MF for different values of the
spectral index  
$\gamma = 0, 1, 2$. We remark that for $\gamma = 1$ the adaptive
filter and the matched filter coincide and for $\gamma=2$ 
the matched filter and the Mexican Hat wavelet are equal.

For the MF the parameters $\beta$ and $\rho$ and the curvature of the
source $y_s$ are given by
\begin{equation}
\beta^2 = \frac{R^2}{\sqrt{m(1 + m)}},~~~ \rho = \sqrt{\frac{m}{1 + m}},~~~
y_s = \rho.
\end{equation}
In Fig.~\ref{parameters} appear the parameters
$\beta\sqrt{\rho} ,\rho$ and $y_s$ as a function of the  
spectral index $\gamma $ for the MF.

\section{Analytical and numerical results} \label{sec:numerical}
We will distinguish the two cases of simple detection and simple
measurement in the applications that follow.
 
\subsection{Simple detection}
\begin{figure}
\centerline{\epsfig{file=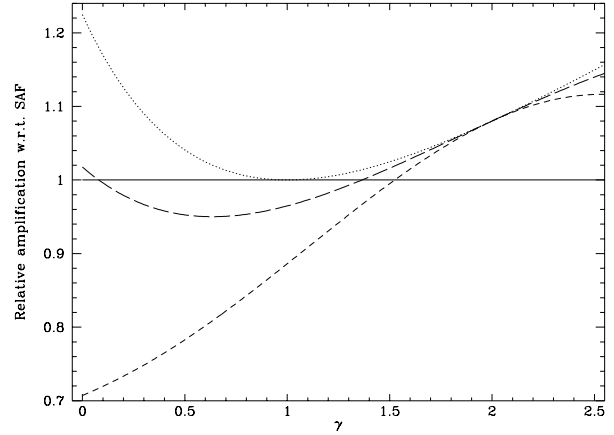,
angle=-90,width=\hsize}} 
\caption{Relative amplification given by the MF (dotted line) and MHW
(short-dashed line) with respect to the SAF (solid line) versus the spectral 
index $\gamma$. For comparison, the amplification given by the MHW$_o$ 
(long-dashed line) is also shown. The MF gives always a higher gain
because it is constructed imposing maximum amplification of the source.}
\label{amplification}
\end{figure}
In this case, we try to detect sources of known amplitude $A$. We
note that the same amplitude $A$ translates into different thresholds
$\nu_{s,i}=A/\sigma_i$ (where $i$ refers to the SAF, MF or MHW) in the
filtered map for the three considered filters, since they have different
amplifications. The relation between the values of $\nu_s$ for each
filter can be easily obtained taking into account their relative
amplification, which is given in Fig.~\ref{amplification}. In the
following comparison, we will consider the same
amplitude of the source for all the filters and its value will be
given as a function of the dispersion of the map filtered by the
$SAF$. 
 
Using eqns. (\ref{pbyp},\ref{lik_nus}) one obtains for
the likelihood $L(\nu,\kappa|\nu_s )$ 
\begin{equation}   
L(\nu ,\kappa |\nu_s ) = e^{\nu_s \varphi - \frac{\nu_s^2}{2}\mu }
{\left[1 + B\left(\frac{\nu_s y_s}{\sqrt{2}}\right)\right]}^{-1},
\label{lik_simpled}
\end{equation}
\begin{equation} 
\varphi (\nu ,\kappa )\equiv \frac{1 - \rho y_s}{1 - \rho^2}\nu +
\frac{y_s - \rho }{1 - \rho^2} 
\kappa,\ \ \ 
\mu \equiv \frac{{(1 - \rho y_s)}^2 }{1 - \rho^2}.
\label{phiymu}
\end{equation}
The region of acceptance $R_*$ is defined by $L\geq L_*$
or equivalently  
$\varphi \geq \varphi_* (\nu_s )$ (we assume $\nu_s \geq 0$) with
\begin{equation}
\varphi_*=\frac{1}{\nu_s}\ln\left[L_*\left(1+B\left(\frac{\nu_s
y_s}{\sqrt{2}}\right)\right)\right] + \frac{\nu_s}{2}\mu
\end{equation}
Clearly, the last constraint can be rewritten
\begin{equation}  
R_*: \ \ \nu \geq \nu_* (\kappa ; \varphi_* ), \ \ \ 
\nu_* \equiv \frac{\rho - y_s}{1 - \rho y_s}\kappa + \frac{1 -
\rho^2}{1 - \rho y_s} \varphi_* .
\end{equation}
\begin{figure}
\centerline{\epsfig{file=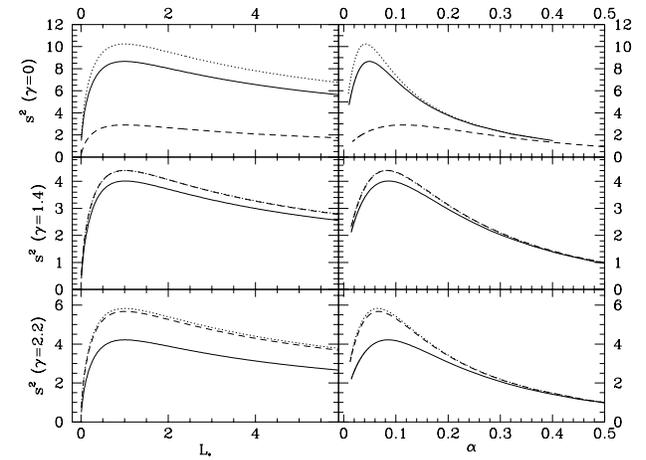,
angle=-90,width=\hsize}} 
\caption{Value of the significance versus $L_*$ and $\alpha$ for
$\gamma=0$ (top), 1.4 (middle) and 2.2 (bottom) for the simple
detection case with a source amplitude corresponding to
$\nu_{s,{\rm SAF}}=3$. The SAF, MF and MHW are given by the solid,
dotted and dashed lines respectively. Note that the maximum 
of $s^2$ is obtained in all cases at $L_*\simeq 1$.}
\label{significance}
\end{figure}
In order to find the acceptance region, we need to maximize the
significance $s^2$ with respect to $\alpha$ (or equivalently $L_*$) 
for each filter. Fig.~\ref{significance} shows the curve of $s^2$
versus $\alpha$ and versus $L_*$ for the three filters with
$\nu_{s,{\rm SAF}}=3$ and $\gamma=0,1.4,2.2$. 
It is remarkable that 
for all the cases the maximum of $s^2$ is found at $L\simeq1$ within 
a few per cent. This is also true independently of the source
amplitude (for the studied range of $\nu_{s,{\rm SAF}}$ from 1 to 5). 
Therefore the chosen criterion indicates that we are in the best
conditions to discriminate between the two hypothesis by using a
simple acceptance region: the
candidates are accepted as detections if the probability of having
background plus source is greater or equal than having only background.
The value of $\alpha$ that maximizes the significance depends mainly on
the amplitude of the source and ranges from 
$\sim 0.3$ for $\nu_{s,{\rm SAF}}=1$ to
$\sim 0.01$ for the highest amplitude considered,
$\nu_{s,{\rm SAF}}=5$. This simply reflects the fact that the higher
the amplitude of the source, the lower the number of spurious
detections and vice versa.
Taking into account the previous results we have calculated the
acceptance region using $L_*=1$ for all the cases studied in this
section.

\begin{figure}
\centerline{\epsfig{file=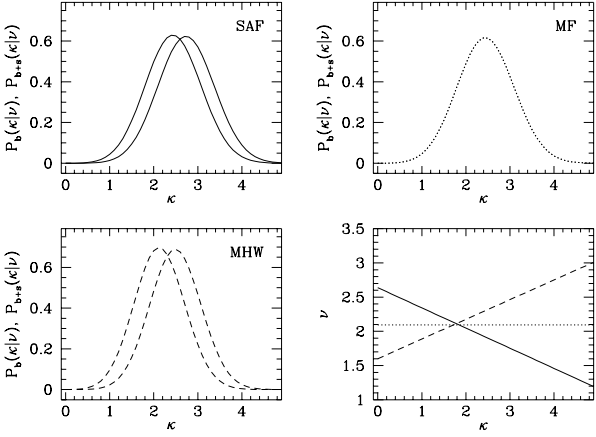,
angle=-90,width=\hsize}}
\caption{The curvature distribution given a threshold
$\nu=\nu_s$ of maxima of the background (thin lines)
and of the background plus source (thick lines) 
filtered with the SAF, MF and MHW are shown in the top and
bottom left panels. The amplitude of the source corresponds to a
threshold $\nu_{s,{\rm SAF}}=3$ and the spectral index of the
background is $\gamma=1.4$.
The bottom right panel shows the acceptance region for the same
case: those maxima with $\nu$ and $\kappa$ above the line are accepted 
as sources and those below are rejected. Solid, dotted and dashed
lines correspond to the SAF, MF and MHW respectively.}
\label{px}
\end{figure}
In the bottom right panel of Fig.~\ref{px} the acceptance region for
the case $\gamma=1.4$ and $\nu_{s,{\rm SAF}}=3$ is given for the
different filters. Those maxima with curvature and amplitude above the 
line are accepted as point sources, while those below are rejected.
The slopes of the lines are easily understood by looking at the other
three panels of the figure, that show
the curvature distribution given a threshold
$\nu=\nu_s$ of maxima of the filtered background (thin line) and of
the background plus source (thick line). For the MF both distributions  
are identical, i.e., we can not differentiate between true and
spurious detections based on their curvature. Therefore, the
acceptance region is fixed by the amplification of the filter: if the
maximum is found above a certain threshold it is accepted,
otherwise is rejected. For the SAF, the distribution of curvature is
shifted to higher values when a source is present. Thus, 
maxima with larger curvature are accepted at lower thresholds than
those with smaller curvature, since the latter are more likely to be
produced by the background. On the contrary, when the field is
filtered with the MHW, the maxima produced by the source tend to have a smaller
curvature than those of the background and the slope of the curve changes
with respect to the SAF case.

Once $R_*$ has been obtained, we can calculate the density number of spurious
sources (corresponding to background fluctuations) $n_b^*$ 
and real detections $n^*$ by integrating
eqns. (\ref{nbackground},\ref{nsource}) in the region of acceptance:
\begin{eqnarray} 
n_b^* &=& \frac{n_b}{2}\left[ {\rm erfc}\left( \frac{\sqrt{1-\rho^2
\varphi_*}}{\sqrt{2}(1-\rho y_s)}\right) \right. \nonumber \\
&& \left. +\sqrt{2}a y_s {\rm e}^{-a^2
\varphi_*^2} {\rm erfc}\left( -\frac{\sqrt{1-\rho^2}}{1-\rho y_s} y_s
a \varphi_* \right)\right], \nonumber \\
a&=&\sqrt{\frac{1-\rho^2}{2(1-2\rho y_s+y_s^2)}},
\label{naceptb}
\end{eqnarray}
\begin{eqnarray}
n^* &=& \frac{n_b}{2}\int_0^\infty d\kappa \kappa e^{-\frac{1}{2}
(\kappa-y_s \nu_s)^2}  
{\rm erfc}(z), \nonumber \\
z&=&\frac{\sqrt{1-\rho^2}}{\sqrt{2}(1-\rho y_s)}\left[\varphi_* -y_s \kappa
-\nu_s \frac{(1-\rho y_s)^2}{1-\rho^2} \right]
\label{nacept}
\end{eqnarray}
\begin{figure*}
\centerline{\epsfig{file=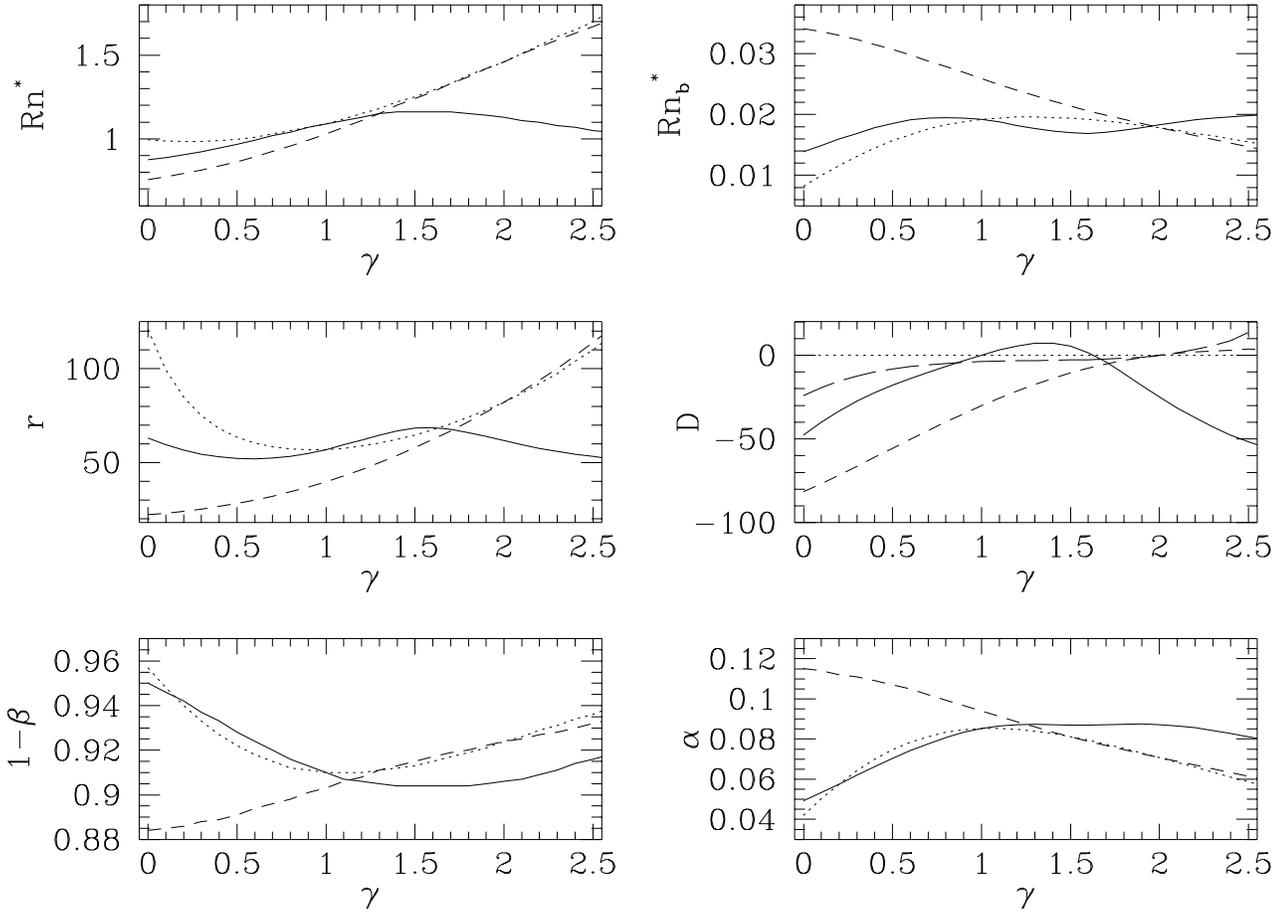,
angle=-90,width=\hsize}}
\caption{Simple detection results for a source amplitude corresponding 
to $\nu_{s,{\rm SAF}}=3$ and an acceptance region obtained fixing
$L_*=1$. Solid, dotted and short-dashed lines refer to the SAF, MF and MHW
respectively. The different panels correspond to the density number of 
detections $n^*$ (top left), the density number of spurious sources
$n_b^*$ (top right), 
the ratio $r$ of true to spurious detections (middle left), the relative
difference $D$ in the ratio with respect to the MF in per cent
(middle right), the probability of identifying correctly a detection, 
given a maximum in the position of the source $1-\beta$
(bottom left) and the probability of misidentifying a peak of
the background as a source $\alpha$ (bottom right). For comparison,
the relative difference in the ratio obtained for the MHW$_o$
(long-dashed line) is also given in the middle right
panel.}
\label{simple_nu3}
\end{figure*}
\begin{figure}
\centerline{\epsfig{file=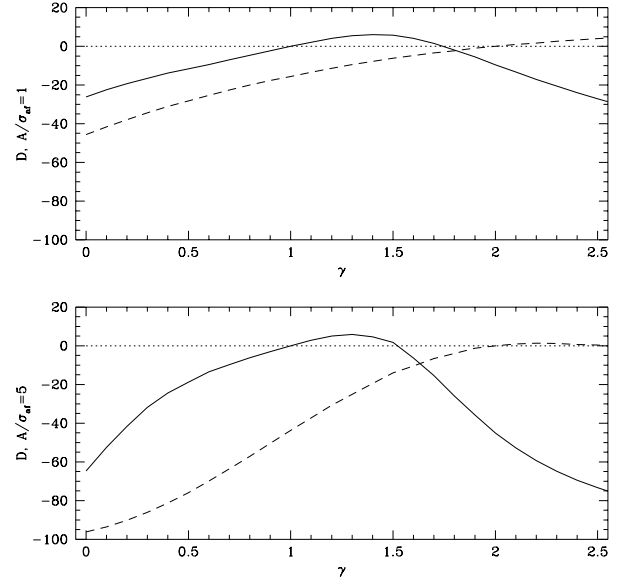, angle=0,width=\hsize}}
\caption{Relative difference in the true to spurious ratio of the SAF
(solid line) and MHW (dashed line) with respect to the MF (dotted
line) in per cent. Top and bottom panels correspond to a source
amplitude of $\nu_{s,{\rm SAF}}=1$ and $\nu_{s,{\rm SAF}}=5$ respectively. 
The acceptance region has been obtained in both cases using $L_*=1$.}
\label{simple_nu1_nu5}
\end{figure}
The number densities $n_b^*, n^*$, the ratio $r=n^*/n_b^*$, $\alpha$ and 
$1-\beta$ have been plotted in Fig.~\ref{simple_nu3} versus the
spectral index $\gamma$ 
for a source amplitude corresponding to $\nu_{s,{\rm SAF}}=3$ for the
SAF, MF and MHW. 
In order to compare the performance of the different filters, we have also
plotted the relative difference $D$ of the detection ratio with
respect to the MF, which is defined as
\begin{equation} \label{eq:D}
D=\frac{r_i-r_{\rm MF}}{r_{\rm MF}}\times 100
\end{equation}
where the subindex $i$ refers to the different filters. Therefore a
positive value of $D$ indicates that the corresponding filter has a
better detection ratio than the MF. Three different regions can be
seen in relation to the performance of the filters. For $\gamma \simeq 
0-1$ the MF clearly outperforms the SAF and the MHW. The SAF works
better than the other two filters in the range $\gamma \simeq
1-1.6$. Finally, at the highest values of $\gamma$ the MHW has the best
performance of the three considered filters, although it is only
slightly better than the MF. This behaviour is qualitatively similar
for other source 
amplitudes, although the performance of the MF in comparison with the
other filters is improved for bigger amplitudes and gets worse for
smaller ones (see Fig.~\ref{simple_nu1_nu5}). 

Although a detailed comparison with the MHW$_o$ is
beyond the scope of this work, we have also plotted the detection
ratio relative to the MF obtained by the MHW$_o$ for
the case $\nu_{s,{\rm SAF}}=3$ using the region of acceptance defined
by $L\ge 1$ (see middle right panel of Fig.~\ref{simple_nu3}). The MHW$_o$ 
clearly outperforms the MHW at the scale of the
source in the considered range of spectral indexes. In fact, this
filter gives the best detection ratio for high values of $\gamma$.

\subsection{Simple measurement}
In this case, the likelihood ratio is given by
\begin{equation}
L(\nu ,\kappa ) = \frac{1}{\nu_c}\int_0^{\nu_c}d\nu_s\,e^{\nu_s
\varphi - \frac{\nu_s^2}{2}\mu } 
{\left[1 + B\left(\frac{\nu_s y_s}{\sqrt{2}}\right)\right]}^{-1}.
\end{equation}
\begin{figure}
\centerline{\epsfig{file=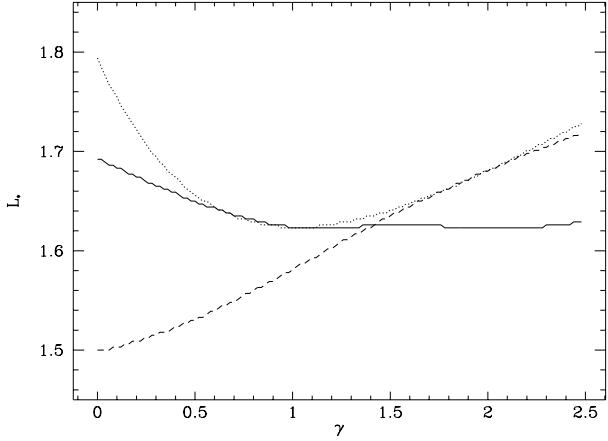,
angle=-90,width=\hsize}} 
\caption{Value of $L_*$ that maximizes the significance versus the
spectral index for the simple measurement case. Solid, dotted and
dashed lines correspond to the SAF, MF and MHW respectively.}
\label{lmax}
\end{figure}
\begin{figure}
\centerline{\epsfig{file=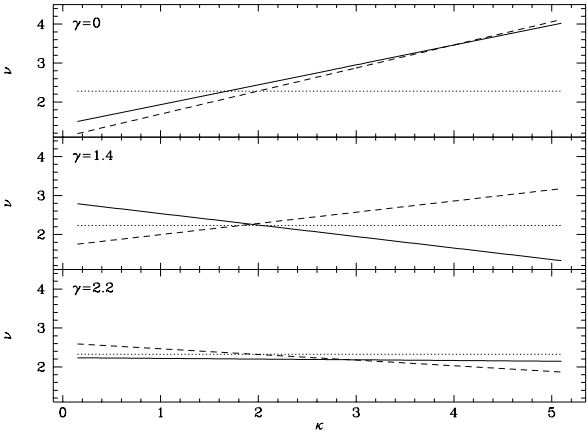,
angle=-90,width=\hsize}} 
\caption{Acceptance region for the simple measurement case for three
different spectral indexes: 0 (top), 1.4 (middle) and 2.2
(bottom). Different lines correspond to the three considered filters:
SAF (solid), MF (dotted) and MHW (dashed).}
\label{reg_intnus}
\end{figure}
As in the previous case, we need to maximize the significance with
respect to $L_*$ in order to get the acceptance region $L\ge L_*$. The
values of $L_*$ 
that maximize the significance are plotted versus the spectral index in
Fig.~\ref{lmax} for $\nu_{c,{\rm SAF}}=5 $. In this case, the
values of $L_*$ are in  
the range of $\simeq 1.5$ to $1.8$ depending on the filter and spectral
index. It also depends on the assumed value of $\nu_c$, higher values
of this parameter lead to larger values of $L_*$. This has an
effect on the values of $L_*$ for the different filters, since the
same amplitude of the source translates into different thresholds in the 
filtered maps and therefore we have a different $\nu_c$ for each
filter. Thus the relative difference in the value of $L_*$ 
between filters seen on Fig.~\ref{lmax} are closely related to the
relative amplification (see Fig.~\ref{amplification}). 
The corresponding acceptance regions for $\gamma=0,1.4,2.2$ are given
in Fig.~\ref{reg_intnus}.

The density number of spurious sources (corresponding to background 
fluctuations) $n_b^*$ and real detections $n^*$ are obtained
by integration of $n_b(\nu ,\kappa )$ and $n(\nu ,\kappa )\equiv 
\frac{1}{\nu_c}\int_0^{\nu_c}d\nu_s\,n(\nu,\kappa |\nu_s)$ in the
region of acceptance $R_*$ respectively. 

\begin{figure*}
\centerline{\epsfig{file=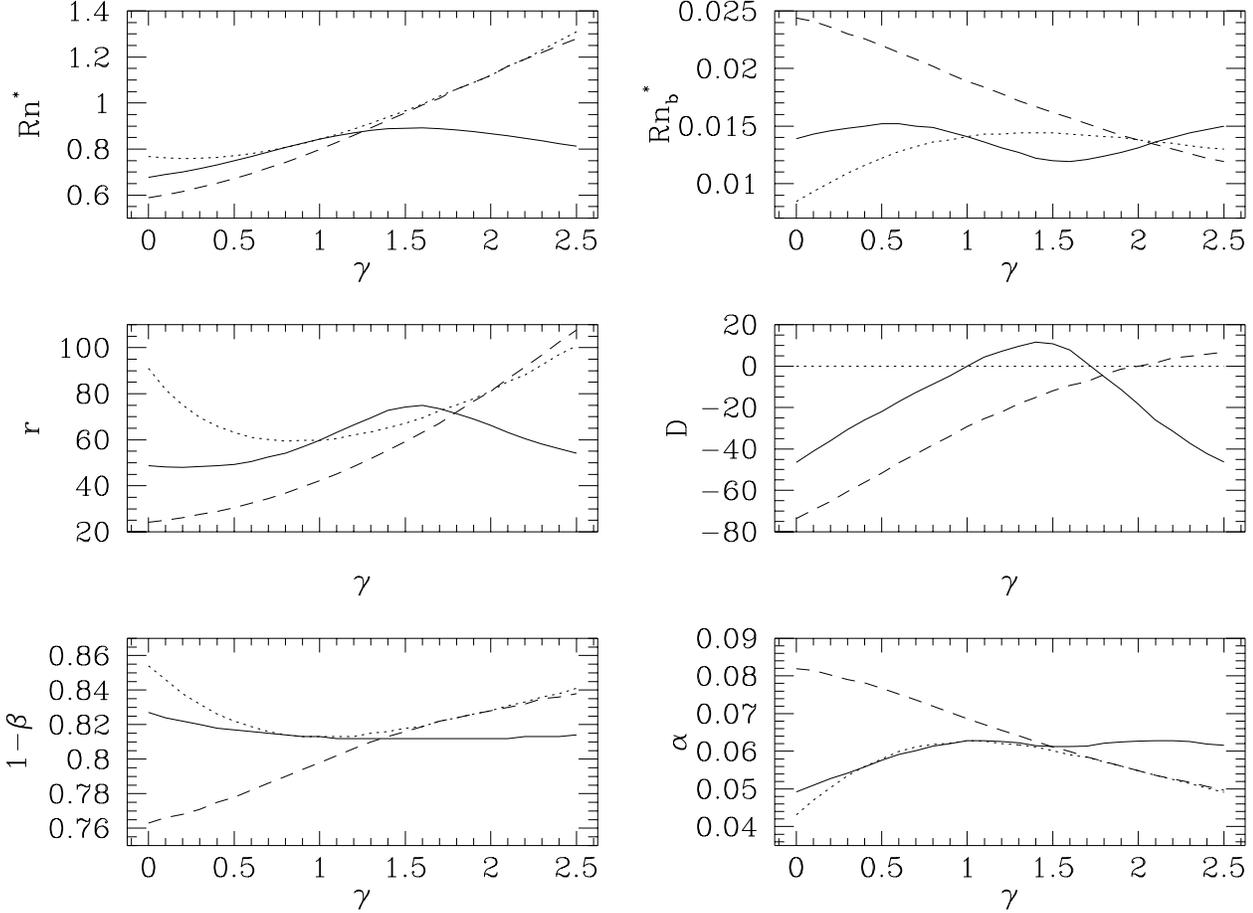,
angle=-90,width=\hsize}} 
\caption{Results for the simple measurement case. The acceptance
region has been obtained maximizing the significance with respect to
$L_*$. The different lines
and panels are the same as those in figure~\ref{simple_nu3}.}
\label{simplem}
\end{figure*}
The quantities $n_b^*$, $n^*$, $n^*/n_b^*$, D, $\alpha$ and $1-\beta$
are given in Fig.~\ref{simplem} as a function of the spectral
index. The acceptance region has been obtained using the value of $L_*$
(given in Fig.~\ref{lmax}) obtained from the maximization of the
significance in each case.
We see that the situation is qualitatively similar to the simple
detection case regarding the relative difference in the detection ratio.
Again the MF outperforms the other filters at low $\gamma$, the SAF has
the best behaviour at intermediate values whereas the MHW
gives a higher ratio at large $\gamma$.

On the other hand, a minimization of the likelihood ratio $L(\nu
,\kappa |\nu_s )$, given by eqn.~(\ref{lik_simpled}), allows 
an estimation of the amplitude of the source ${\hat{\nu}}_s$ to be obtained
as a function of the data $(\nu ,\kappa )$. The estimation of the
amplitude is then given as the solution of the equation
\begin{equation} \label{eq:estimation_of_the_amplitude}
(y^2_s + \mu) \nu_s - \varphi + \frac{1}{\nu_s}\frac{B
\left(\frac{\nu_s y_s}{\sqrt{2}} \right)
}{1 + B\left(\frac{\nu_s y_s}{\sqrt{2}} \right)} = 0,
\end{equation}    
where the function $B$ is given by
eqn.~(\ref{totalnsource}) and $\mu$ and $\varphi$ by eqn.~(\ref{phiymu}).
A confidence level for the estimation of the amplitude of the source
can also be obtained from eqn.~(\ref{lik_simpled}).

\section{NUMERICAL SIMULATIONS: RESULTS} \label{sec:simulations}

Let us consider how the ideas presented in the previous sections
apply to real cases. 
In order to do this, we simulated a set of one-dimensional 
`images' 
containing a Gaussian background characterised 
by a power spectrum $P(q) \propto q^{-\gamma}$.
Two different experiments have been carried out.
In the first set of simulations Gaussian sources of known
width and amplitude were introduced. This case corresponds with 
the \emph{simple detection} scheme presented in 
section~\ref{sec:simple_detection}, in which the knowledge of
the amplitude of the source (or, equivalently, its normalized
amplitude $\nu_s$) can be used to determine the region
of acceptance.
In the second set of simulations the Gaussian sources had
amplitudes that were drawn from a
uniform distribution 
between a minimum and a maximum value. These simulations were
used to test the \emph{simple measurement} scenario, in which
the question is not only whether the source is detected or
not, but also if its amplitude can be estimated as well.

\subsection{Simulations in the simple detection scenario}   \label{sec:simulations_sd}

In order to test the performance of different filters in the
simple detection scenario, we simulated a set of `images' with
$N=1024$ pixels each 
and a Gaussian background characterised 
by a power spectrum $P(q) \propto q^{-\gamma}$.
For each simulated image,
a source with a Gaussian profile of width $FWHM=5$ pixels and known
amplitude $A_o$ was placed at the central 
pixel. Then, the image was filtered with a matched filter, a
scale-adaptive filter and a Mexican Hat wavelet. 
The matched and scale-adaptive
filters were chosen taking into account the source width and the 
background index as in eqns. (\ref{eq:saf}) and (\ref{eq:mf}),
whereas the Mexican Hat had the same width than the Gaussian profile
of the source. The normalized amplitude
of the filtered source is $\nu_{s,X}=A_o/\sigma_X$, where $X$ refers
to the filter in consideration (SAF, MF or MHW).
After filtering, we look for a maximum at the position
of the source we introduced. 
Here we implicitly assume that we know the exact position of the source,
which is true for our simulations but will not be the case in a
realistic observation. Fixing the position of the source allows us
to clearly illustrate the behaviour of the optimal statistics described
above. A short comment on position uncertainties will be presented in
section \label{sec:position}. If there exists such a maximum,
we measure its normalized amplitude $\nu$ and curvature $\kappa$.
With these quantities and the previously known $\nu_s$ it is
easy to determine whether the maximum lies inside a certain region of
acceptance $R_*$ or not.
We adopt the criterion C1 introduced in section~\ref{sec:simple_detection}:
$R_* : L(\nu,\kappa) \geq L_*$, and choose $L_*=1$ (for a
justification for this particular value see section~\ref{sec:numerical}).
If the maximum is in the region of acceptance, we
count it as a \emph{valid detection} and, if not, as a \emph{rejection}
(a false dismissal). The number of rejections is directly related
with the probability $\beta$ of eqn.~(\ref{alfaybeta}).

On the other hand, let us consider a simulation with the same
background, but without any source, and  filter it
in the same way as before. Now, let us 
repeat the same procedure than before: we look for a local maximum
at the central pixel of the `image' and,
in case it exists, we apply the selection mechanism, that is,
make the (wrong) hypothesis that a source  
of normalized amplitude $\nu_s$ is present in the simulation, and
determine the region of acceptance as before. If the maximum is not inside
the region of acceptance, it has been safely rejected. On
the contrary, if it lies inside the region of acceptance, the
noise will be interpreted as a signal and a false detection
will occur. The number of false detections is directly related
with the probability $\alpha$ of eqn.~(\ref{alfaybeta}).

In order to test the behaviour of the
probabilities of detecting a signal, rejecting a signal
and obtaining a false detection
 as functions
of the source amplitude $A_o$ and the filter, we tested 10
different values of $A_o$.
Three different background regimes were tested: background
index $\gamma=0$, corresponding to a case
in which the MF is expected to detect with better reliability
than the other two filters, $\gamma=1.4$, corresponding to a case
in which the SAF is the most reliable,
and $\gamma=2.2$, where the most reliable filter
is the MHW (see section~\ref{sec:numerical}).
 The lower and higher amplitude limits
$A_o^{min}$ and $A_o^{max}$
were chosen so that after filtering with the SAF 
the normalized amplitudes are $\nu_{s,SAF}^{min} \simeq 1$
and $\nu_{s,SAF}^{max} \simeq 5$.
For each value of $A_o$ we performed 10000 simulations containing
a source in the central pixel and other 10000 simulations containing
only background. 
Each simulation was filtered using the SAF, the MF and
the MHW. 
For each pair of filtered simulations (one with source and one
without it) the following quantities were 
recorded: the number of times that a source was properly 
detected $N_d$, the number of times a source produced a
maximum but it was rejected by the selection criterion $N_r$ and
the number of spurious detections (due to the background) that happened in the
simulations without source, $N_b$. The total number
of maxima found when a source is present is $N_t=N_d+N_r$.
Another interesting quantity is the ratio $r=N_d/N_b$, that
gives us an idea of the practical reliability of the
detection.

\subsubsection{The case $\gamma=0$}

\begin{figure*}
\includegraphics[angle=270, width=17cm]{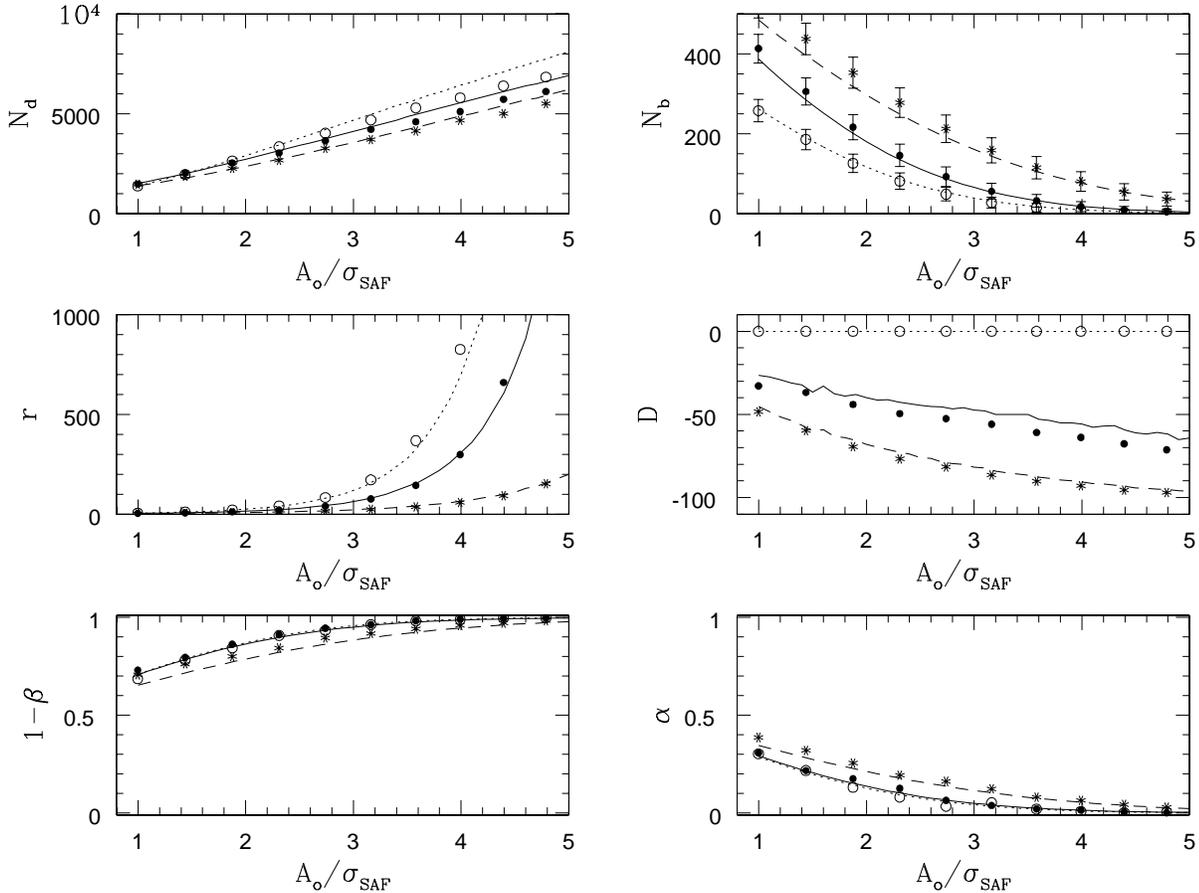}
\caption{Simple detection results for the case $\gamma=0$. Solid lines and 
filled circles make reference to results obtained with the scale-adaptive
filter. Dotted lines and open circles make reference to results obtained 
with the
matched filter.
Dashed lines and asterisks make reference to results obtained 
with the Mexican Hat wavelet.
The different panels are explained in the text.}
\label{fig:sd_gamma0}
\end{figure*}

The results of the simulations with $\gamma=0$ are shown 
in figure~\ref{fig:sd_gamma0}.
The results of the simulations, indicated with points, are compared
with the theoretical curves obtained in section~\ref{sec:numerical}. 
Solid lines and 
filled circles make reference to the scale-adaptive
filter. Dotted lines and open circles make reference to the
matched filter.
Dashed lines and asterisks make reference to the Mexican Hat wavelet.

The first panel on the top and the left of figure~\ref{fig:sd_gamma0}
 shows the 
number of detections $N_d$ (local maxima due to sources that are inside
the region of acceptance) in the 10000 simulations for the three 
filters. $N_d$ is directly related with the number density of maxima $n^*$,
being the relationship between the two quantities 
dependent on the sizes of the pixel and the filter scale $R$.
Since the matched filter has the greatest gain, it
gives the highest number of detections. 
The number of detections found in the simulations is lower than
the expected value. This result is not unexpected due to
the following reason: the theoretical expected number of detections
is calculated by multiplying the number density of maxima in the
presence of a local source $n^*$ 
by the number of simulations (10000 in this case) and the pixel size
(in units of the filter width $R$). This product can, eventually, give
more than one theoretical detection per simulation and pixel, if
the pixel size and $\nu_s$ are big enough. 
Obviously, in a real simulation
one can only find zero or one maxima per pixel. 
This effect is more conspicuous for high $\nu_s$ and,
as we will see later, for higher $\gamma$ values, 
producing apparent paradoxes such as the prediction, in some cases, of
a number of detections greater than 
the number of simulations
we have. 
Therefore, the
theoretical curve must be considered in the best case only 
as a upper limit for the real number of detections. 

The panel on the top and the right of figure~\ref{fig:sd_gamma0}
shows the number of spurious detections $N_b$ found in the 10000 simulations
versus the expected value. 
Again, $N_b$ is directly proportional to the number density of
maxima due to the background $n^*_b$.
The dots were obtained by counting
the number of accepted background maxima considering all the
pixels in absence of local source and then computing the average 
number of spurious detections per pixel, and the statistical $1\sigma$
error bars of such average. The agreement between the expected values
and the results from the simulations is remarkable. The matched
filter gives the lowest gross number of spurious detections.

The left and the right lower panels in figure~\ref{fig:sd_gamma0}
show the ratio between the number of background maxima interpreted as source ($N_b$) and
the total number of maxima due to the background, 
and the ratio between source
detections ($N_d$)  and the total number of maxima in presence of a source 
($N_t=N_d+N_r$), respectively. 
The first ratio gives us an estimation of the false alarm probability $\alpha$ whereas
the second one gives the power of the detection $1-\beta$.
The results obtained
with the SAF and the MF are strikingly similar to the expected
values, whereas the concordance is slightly worse in the case of
the MHW. Note that the problem of the theoretical curves mentioned above 
does not affect these ratios because the terms that come from
the number of simulations and the pixel area cancel in the division.

More interesting is the ratio between the number of effective
detections and the number of spurious ones. 
This quantity is shown
in the middle panel on the left of figure~\ref{fig:sd_gamma0}.
Since the proportionality between $N_d$ and $n^*$ is the same
than the one between $N_b$ and $n^*_b$, this ratio is
a good estimator of the ratio $r=n^*/n^*_b$.
The inverse of this ratio gives the percentage of
spurious detections that is found in a number $N$ of 
detected maxima. Again, the agreement between theory and
simulations is excellent. Note that the greatest values of $r=N_d/N_b$
correspond to the matched filter. That means that, if what
we are looking for 
is \emph{reliability} in the detection, the matched filter is
the best choice in this case $\gamma=0$. 

In order to quantify the differences between filters regarding
reliability (in the sense mentioned above),
it is useful the quantity $D$ defined in equation (\ref{eq:D}).
This $D$ can be understood as the per cent relative difference
between filters.
In the middle panel on the right of figure~\ref{fig:sd_gamma0}
it is shown the value of $D$.
The differences between the MF and the other filters are
$\sim 50\%$ for the SAF and $\sim 80\%$ for the MHW 
at intermediate $\nu_s$ values.

Gaussian backgrounds with $\gamma=0$ correspond to uncorrelated
(white) noise and are very common in many fields of physics. For
instance, the background of many astronomical images is dominated by
uncorrelated Poissonian noise, which is well approximated by this kind
of noise when averaging over many observations.

\subsubsection{The case $\gamma=1.4$}  

\begin{figure*}
\includegraphics[angle=270, width=17cm]{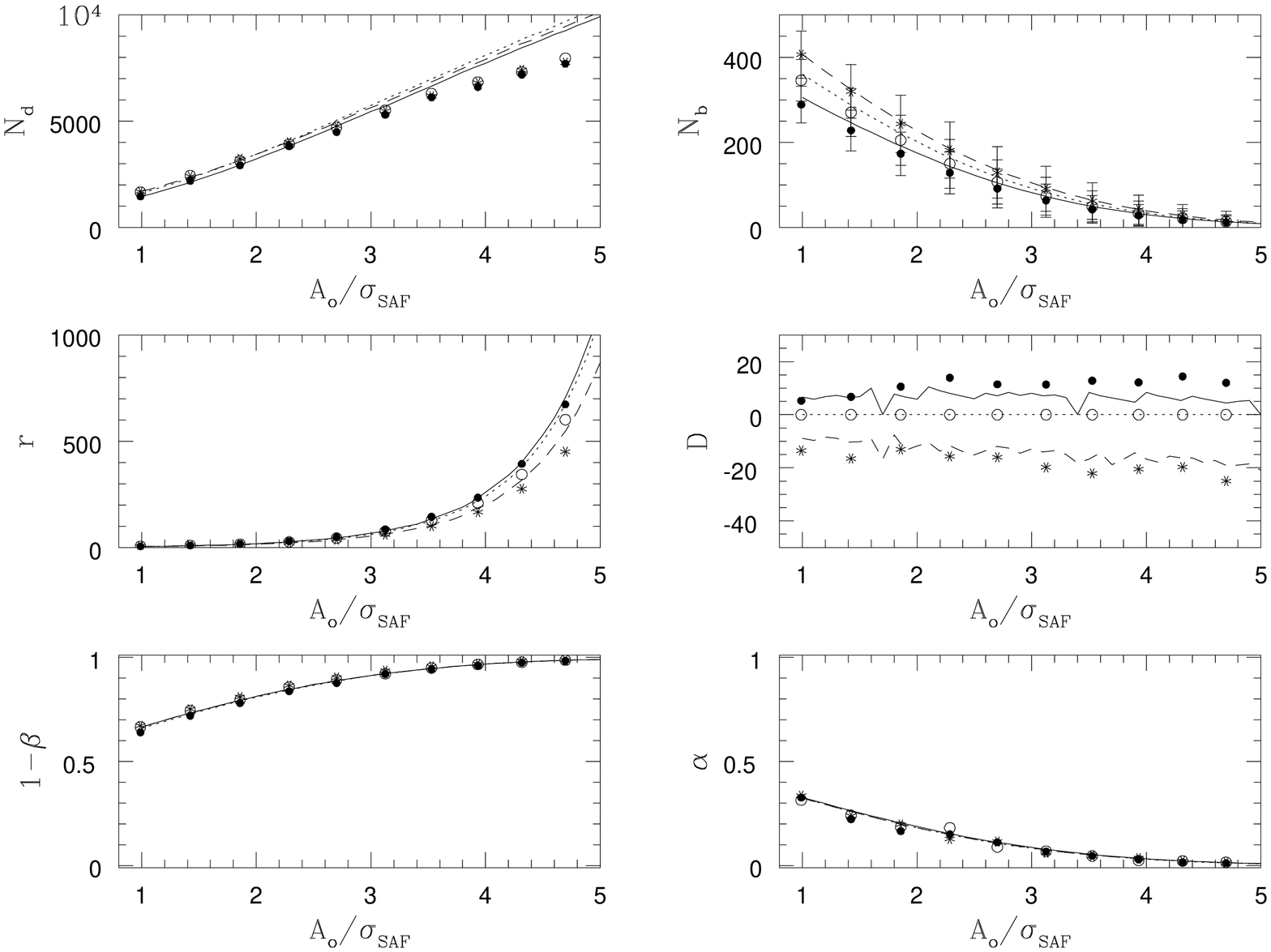}
\caption{Simple detection results for the case $\gamma=1.4$. 
The different lines and panels are the same as those in figure~\ref{fig:sd_gamma0}.}
\label{fig:sd_gamma1p4}
\end{figure*}

The results of the simulations with $\gamma=1.4$ are shown 
in figure~\ref{fig:sd_gamma1p4}.
As in the case before, 
the results from the simulations are compared with 
the  theoretical curves obtained in section~\ref{sec:numerical}. 
The different lines and panels are the same as those in figure~\ref{fig:sd_gamma0}.
The agreement between the theoretical expectations and the results
is very good.

The number of detections $N_d$ at a given normalized amplitude $\nu_s$ is
higher for the three filters than in the case $\gamma=0$. The filtering is more
efficient at enhancing sources when $\gamma$ increases due to the fact
that at high $\gamma$ the power of the background concentrates 
in large scale structures that are more easily removed by the filters.

In this case the performance of the three filters is similar
in number of detections as well as in the probabilities $\alpha$ and
$1-\beta$. The SAF, however, produces a lower number of spurious
detections due to the background. Therefore, the ratio $r$ between
authentic and spurious detections is better for the case of the SAF 
(although the MF performs almost as well in this aspect). The relative
difference $D$ between SAF and MF in this case takes values around
$10\%$. The MHW performs significantly worse than the other two filters
in this case.
 
We remark that noise index in the interval $1 \leq \gamma \le 1.6$ are not
rare in some areas of Astronomy. For example, in the scanning of the sky
of many CMB experiments (MAP, Planck, etc) backgrounds with noises in
this range may appear due to the combination of CMB and Galactic foregrounds
with the scanning $1/f$ noise.

\subsubsection{The case $\gamma=2.2$}

\begin{figure*}
\includegraphics[angle=270, width=17cm]{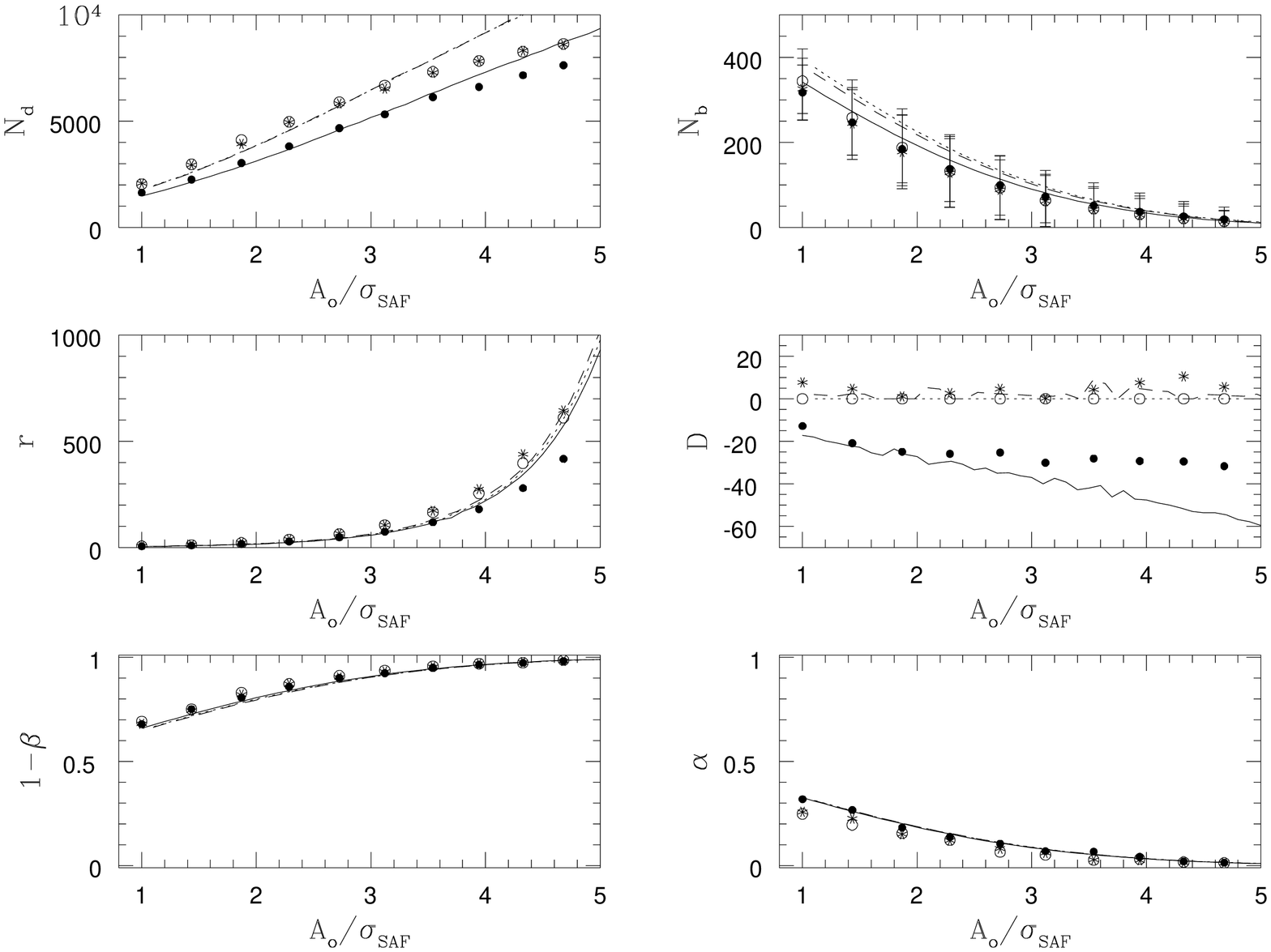}
\caption{Simple detection results for the case $\gamma=2.2$. 
The different lines and panels are the same as those in figure~\ref{fig:sd_gamma0}.}
\label{fig:sd_gamma2p2}
\end{figure*}

The results of the simulations with $\gamma=2.2$ are shown 
in figure~\ref{fig:sd_gamma2p2}.
As in the two cases before, 
the results from the simulations are compared with 
the expectations obtained in section~\ref{sec:numerical}. 
The different lines and panels are the same as those in figure~\ref{fig:sd_gamma0}.

This case corresponds with the region in which the MHW outperforms the other
two in reliability. The matched filter, however, performs almost as well as
the MHW. The relative differences $D$ in the detection ratio are of
a few percent between the MHW and the MF. The SAF performs significantly
worse in this case.

The agreement between the results of the simulations and the
theoretical predictions 
is good again, though not so good as in the previous cases.
The reason is that in practice it is not possible to simulate perfectly 
a $P(q) \propto q^{-\gamma}$ background with $\gamma > 0$ because such 
backgrounds diverge when $q$ tends to $0$. The discrepancy between the
ideal power spectrum and the simulated one gets worse when $\gamma$ increases.

Spectral indices around 2 correspond to smooth backgrounds and
are less common than smaller indices. However, they appear in
certain cases. For instance, microwave observations dominated by dust
emission have $\gamma \sim 2$.

\subsubsection{Determination of the position} \label{sec:position}

In the simulations presented above, our aim was reproducing the
theoretical scheme using simulations. Therefore,
we have focused only on what
happens in one pixel of the image, the pixel in which the source
is located. However, in most real situations, the position of the
source is not known and we will need to consider all the pixels of the
image. 

Note that we are assuming that the position of the maximum is a good
estimator of the position of the source. In practice, this is not
necessaraly true. In particular, the maximum can be shifted from the
position of the source due to background fluctuations, although
the probability of finding  a peak due to source plus background 
decreases quickly as the distance to the source position increases. 
We have calculated the probability of a peak appearing at a certain
distance $d_p$ (in pixel units) from the position of the simulated source 
for the case $\gamma=1.4$. We expect the results to be qualitatively
similar for the other cases.
At $\nu_s=5.0$ we found
that the probability of the peak appearing at distance $d_p=1$ is 
approximately $0.2$ for the three considered filters. 
At $\nu_s=3.0$ this probability increases to $0.4$, whereas at
$\nu_s=1.0$ it increases to $0.5$. When we consider 
$d_p=2$ pixels we found these probabilities to be
practicaly zero for cases $\nu_s=5$ and $\nu_s=3$, and 0.2 for $\nu_s=1$.
For $d_p \ge 3$  the probability is almost zero for
all cases. 
Therefore, in a realistic situation where we need to consider all the
pixels of the image, 
and in particular the neighbouring pixels of the source, the
number of detections will slightly increase (in a very similar
manner for all the considered filters). 
Regarding the number of spureous sources, they do not depend on the
pixel position. Thus the spureous detections will increase linearly with
the considered number of pixels. This behaviour is again the same for
all the filters.
Therefore the main conclusion is that while the global performance of the
filters will be worse in 
a realistic case in which the position of the sources is not known, 
the relative performance of the filters will still be the same 
as that described in the previous sections. 

Note that in our simulations the FWHM of
the beam is 5 times the pixel size and that basically all maxima produced by
a point source fall within the beam size. Therefore, by simply using the
position of the maximum as an estimator of the position of the source,
this last quantity can be determined with a precision given by the
beam size of the experiment.
More ellaborated techniques can be used to improve the determination
of the position. For instance one could use the average 
of the  points at half the peak intensity along the slopes of the image,
or fitting these points to a certain profile. This is out of the 
scope of the present paper and remains for a future work.

\subsection{Simulations in the simple measurement scenario}
\label{sec:simulations_se} 

Now consider that the true amplitude of the source is unknown.
This is the most frequent case in practise.
In such a case, one must both detect and estimate the
amplitude of the sources that are supposed to be
embedded in the data. 
This problem can be 
confronted by adopting the methodology considered at the 
end of section~\ref{sec:simple_measurement}: in absence of any
a priori information about the p.d.f. $p(A)$ we
assume uniformity in the interval $[0,\nu_c ]$,
where $\nu_c$ is some cut threshold over which it is
not expected to find sources. 
An approximate value of $\nu_c$ can be 
obtained directly from the data under consideration simply
by adopting $\nu_c \simeq \left( |I_{max}|+|I_{min}| \right)/\sigma$, 
where $I_{max}$
and $I_{min}$ are the maximum and minimum data values, respectively.
In practise, it is very unlikely that a source with normalized amplitude
greater than the previous value is present in the data.
If the p.d.f. $p(A)$ is known, that information can be used instead 
of the previous uniform distribution. 
As explained in section~\ref{sec:simple_measurement}, using the
p.d.f. $p(A)$ it is possible to determine a region of acceptance for
the detections and the normalized likelihood $L(\nu, \kappa | \nu_s)$.
By means of the normalized likelihood one can estimate the most
probable value  and confidence
limits (`error bars') for the amplitude $A$ of any detected peak.

To test this methodology a set of simulations similar to
the ones described in section~\ref{sec:simulations_sd}
were performed. Half of the simulations had a Gaussian source 
of $FWHM=5$ pixels located at the central pixel, whereas
the other half had only noisy background. Now we
try to detect sources in both kind of simulations without
knowing a priori the amplitude of the (hypothetical) source.
In each case, a region of acceptance is determined as
in eqns. (\ref{eq:likelihood_se},~\ref{eq:acceptance_se}). If a peak is found
and it lies inside the region of acceptance, its most
probable amplitude is estimated 
by looking for the maximum value of the 
normalized likelihood $L(\nu, \kappa | \nu_s)$
and the $68\%$ confidence limits (`error bars') 
are calculated using the normalized likelihood  $L(\nu, \kappa | \nu_s)$.
The number of `detections' found in the simulations without
source indicates the probability of spurious detection
in this scenario.

The amplitude of the simulated sources
takes values between 0 and a maximum value $A_{max}$ so that
$A_{max}/\langle \sigma_{SAF} \rangle = 5$, being 
$\langle \sigma_{SAF} \rangle $
the average value of $\sigma_0$ over 1000 simulations 
filtered with the SAF. We take this value as the value of $\nu_c$ 
for the calculations. For the sake of simplicity
and without any loss of generality, 
we consider 25 amplitude bins in the interval $[0,\nu_c]$
instead of a continuous sampling. A number of 5000
simulations with source and 5000 simulations without
source were performed for every amplitude bin.
As in the previous section, three cases were considered:
background indexes $\gamma=0$, $\gamma=1.4$ and
$\gamma=2.2$.

\subsubsection{The case $\gamma=0$}

In the simple measurement scenario we need to take two aspects into account.
On the one hand, there is the pure detection aspect of the 
process, that is, how many sources are detected, and how many 
spurious detections due to background are accepted by our decision criterion.
On the other hand, there is the issue of how well the 
parameters of the sources (i.e. their amplitude) are estimated.

The three first rows of table~\ref{tab:se_results} show
the relevant results for the simulations with $\gamma=0$ and the
three filters. The theoretical expectation for the number of simulations
we performed is shown in parenthesis. The number of detections is slightly
lower in the simulations than in the theoretical prediction, as we have
seen in section~\ref{sec:simulations_sd}, and for the same reasons that
were explained there. The number of spurious detections $N_b$ was slightly
higher than the theoretical expectation. 
The lower number of detections and higher number of spurious sources lead
to ratios $r=N_d/N_r$ that are worse than in the theoretical case. The difference,
however, is never greater than a $20\%$, and the theoretical behaviour 
is preserved: the MF gives the best ratio $r$, followed by the SAF and then
by the MHW. The agreement in the power
of $1-\beta$ of the filters is very good for the three filters.

\begin{figure}
\includegraphics[angle=270, width=\hsize]{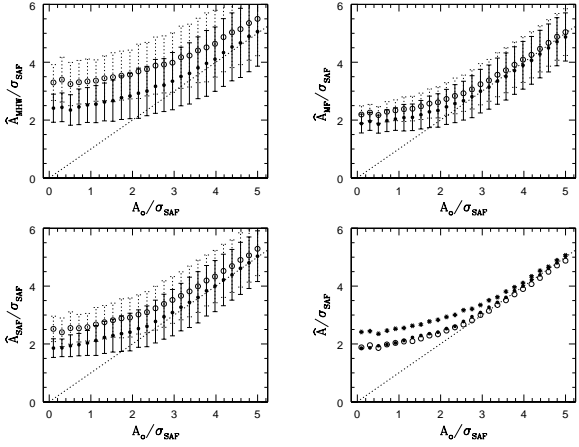}
\caption{Determination of the amplitude in the case of simple measurement and $\gamma=0$.
The estimates using eqn.~(\ref{eq:estimation_of_the_amplitude} (filled circles) and
using the measured intensity (open circles) are shown for the three filters. The lower right panel
shows the comparison between the three 
filters (filled circles for SAF, open circles for MF, asterisks for MHW). }
\label{fig:se_gamma0}
\end{figure}

Regarding the estimation of the amplitudes, the results are shown in
figure~\ref{fig:se_gamma0}. The filled circles show the mean value
of the estimated normalized amplitude as a function of the true
normalized amplitude of the source. The estimation has been
performed using eqn.~(\ref{eq:estimation_of_the_amplitude}).
As a reference, the mean values of the directly measured
amplitudes are shown by the open circles. Clearly, the 
estimation based in the maximization of the likelihood
$L(\nu , \kappa | \nu_s )$ is more accurate than the naive
estimation using directly the measured values of the
maxima. Error bars show the dispersion of the estimates
around the mean value in the simulations (solid line 
for the estimation using eqn.~(\ref{eq:estimation_of_the_amplitude}) 
and dotted line for the naive estimation). 

For high amplitudes the estimation of the amplitude is unbiased
for the three filters. At low amplitudes there is a positive
bias due to a selection effect: the sources are very faint and
only in the cases when a background fluctuation enhances
by chance the source the associated maximum is able to
enter in the acceptance region $R_*$. Using
eqn.~(\ref{eq:estimation_of_the_amplitude}) 
we are able to estimate reasonably well amplitudes as low as
$2.5\sigma_{\rm SAF}$. 

\begin{figure}
\includegraphics[angle=0, width=\hsize]{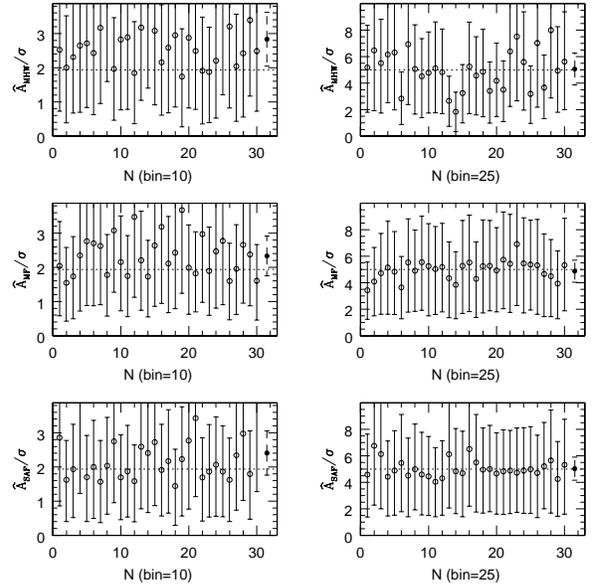}
\caption{Some examples of estimation of the amplitude with its error bars
calculated using eqns. (\ref{eq:estimation_of_the_amplitude}) 
and (\ref{lik_simpled}) for the case $\gamma=0$. 
In each panel 30 randomly chosen detections are shown (open circles) with
their $68 \%$ confidence intervals. The three panels on the left show
detections for the source amplitude bin number 10 (correspondent to $\nu_s \simeq 2$)
whereas the panels on the right show detections for the source amplitude bin
number 25 ($\nu_s \simeq 5$), for the three filters. The horizontal dotted 
line shows the true amplitude of the
sources. As comparison, the mean value of the estimated amplitude
for all the detections in the same bin is shown on the right of each panel
(filled circle) with its $1\sigma$ statistical error bar. 
}
\label{fig:ci_gamma0}
\end{figure}

\begin{table*}
\begin{center}
% \begin{minipage}{170mm}	
 \caption{Results from the simulations in the simple measurement scenario, compared with
the theoretical expectations (in parenthesis). Three background (`noise') regimes
are considered: $\gamma=0$, $\gamma=1.4$ and $\gamma=2.2$. For each value of $\gamma$, three
filters have been studied: the scale-adaptive filter (SAF), the matched filter (MF) and
the Mexican Hat wavelet (MHW). The third column shows the number of
correct detections from
the total of simulations performed (125000). The fourth column shows
the number of 
spurious detections (due to background fluctuations in absence of
sources) per pixel from the total 
number of simulations performed. The fifth column shows the power of the detection (i.e. $\beta$ is
the probability of interpreting signal as noise). The sixth column shows 
the ratio between the number of true and spurious detections, on the basis of
a probability of $50\%$ of presence of a source in the pixel considered.  }
 \label{tab:se_results}
 \begin{tabular}{c |  c c c c c}

\hline
    & Filter & $N_d$ & $N_b$ & $1-\beta$ & $r=N_d/N_b$  \\
\hline

     & SAF &  37863 (39914) & 912  (818)  & 0.84 (0.83) & 41.52 (48.79) \\
$\gamma=0$    & MF  &  41199 (45154) & 498  (496)  & 0.85 (0.85) & 82.73 (91.04) \\
     & MHW &  33742 (34616) & 1715 (1436) & 0.79 (0.76) & 19.67 (24.11) \\

\hline

     & SAF & 46668 (52277)  & 630  (718)  & 0.80 (0.81) & 74.08 (72.81) \\
$\gamma=1.4$ & MF  & 48575 (55279)  & 768  (848)  & 0.81 (0.82) & 63.25 (65.19) \\
     & MHW & 48266 (54455)  & 925  (983)  & 0.81 (0.81) & 52.18 (55.40) \\

\hline

     & SAF & 51396 (49863)  & 744  (824)  & 0.82 (0.81) & 69.08 (60.51) \\
$\gamma=2.2$  & MF  & 60758 (70056)  & 721  (795)  & 0.83 (0.83) & 84.26 (88.12) \\
     & MHW & 60077 (70058)  & 697  (765)  & 0.82 (0.83) & 86.19 (91.58) \\

\hline
 \end{tabular}
% \end{minipage}
 \end{center}
\end{table*}

The lower panel on the right of 
figure~\ref{fig:se_gamma0} show the mean values of the
estimated amplitudes using eqn.~(\ref{eq:estimation_of_the_amplitude}) 
for comparison between the three filters. SAF is denoted by
filled circles, whereas open circles and asterisks stand for MF and
MHW, respectively. The three filters show a very similar 
efficiency at high amplitudes, whereas at low amplitudes
the MF and the SAF work better than the MHW.

As noted in section~\ref{sec:simple_measurement}, using the normalized
likelihood $L(\nu , \kappa | \nu_s )$
it is possible to give a confidence interval for the estimation
of the amplitude. To illustrate this point, 
figure~\ref{fig:ci_gamma0} 
shows the estimated amplitude and the $68\%$ confidence intervals
calculated using eqn.~(\ref{lik_simpled}) for the three filters in two
different cases: when the source is in the limit of detection (left panels
of the figure) and when the source has a good SNR after filtering (right).
Only 30 examples were randomly chosen for each filter and amplitude bin in order to
make the plot readable. For comparison, the mean value and statistical error
bar that appears 
in figure~\ref{fig:se_gamma0} is shown at the right of each plot. The confidence
intervals are significantly larger than the statistical error bars. Hence, they
must be considered as upper limits to the true error. However, in a realistic
case they are the only thing that one can safely say about
the error distribution without having to resort to simulations.

\subsubsection{The case $\gamma=1.4$}

\begin{figure}
\includegraphics[angle=270, width=\hsize]{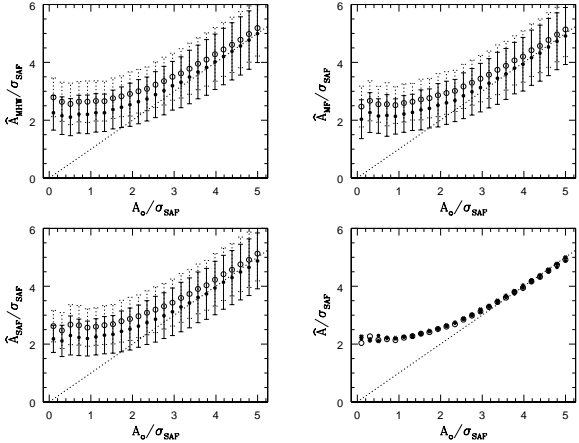}
\caption{Determination of the amplitude in the case of simple
measurement and $\gamma=1.4$. 
The estimates using eqn.~(\ref{eq:estimation_of_the_amplitude}) (filled
circles) and 
using the measured intensity (open circles) are shown for the three
filters. The lower right panel 
shows the comparison between the three filters (filled circles for
SAF, open circles for  
MF, asterisks for MHW). }
\label{fig:se_gamma1p4}
\end{figure}

The rows 4 to 6 in table~\ref{tab:se_results} shows the results for the simple measurement scenario
and $\gamma=1.4$. As expected, the total number of detections $N_d$ has increased
with respect to the case $\gamma=0$, while the number of spurious detections $N_b$ has
decreased.
The relative difference between both numbers and their theoretical expectations 
are below the $13\%$ in all the cases. The lowest number of spurious detections
corresponds to the SAF. 
Moreover, the best ratio $r=N_d/N_b$ corresponds to the SAF as well.
The differences between the predicted $r$ and the ratio given by the simulations is
below the $3\%$ for the SAF and the MF, and below the $6 \%$ in the case of
the MHW. The agreement in the probability $1-\beta$ is very good for the three cases.

Figure~\ref{fig:se_gamma1p4} shows the results of the estimation of the
amplitude in the case $\gamma=1.4$. The meaning of the panels, points and lines
is the same than in figure~\ref{fig:se_gamma0}. Again, the estimation of the amplitude
using eqn.~(\ref{eq:estimation_of_the_amplitude}) is better than the simple estimation
using directly the measured intensity of the maxima. The three filters perform very
similarly in this aspect. As in the previous case, the 
effective limit for a good estimation of the amplitude
is around $\nu_{s,{\rm SAF}} \simeq 2.5$. We would like to remark,
however, that since the filters produce greater amplification in this
case, the absolute 
(no normalised) amplitude one can detect and correctly estimate is in this case
lower than in the case $\gamma=0$.
 
The estimation of confidence intervals works exactly in the same way
as before.

\subsubsection{The case $\gamma=2.2$}

The rows 7 to 9 in table~\ref{tab:se_results} shows the results for the simple measurement scenario
and $\gamma=2.2$. The total number of detections $N_d$ is greater than in previous cases,
that is, the amplification (gain) of the filters is greater than for lower $\gamma$ values.
In the same way,  the number of spurious detections $N_b$ is smaller than before.
The relative difference between both numbers and their theoretical expectations 
are below the $15\%$ in all the cases. As expected, the lowest number of spurious detections
and the best ratio $r$
correspond to the MHW (but the MF performs almost equally well).
Again, there is a remarkable agreement in the probability $1-\beta$ between 
simulations and theoretical expectations.

\begin{figure}
\includegraphics[angle=270, width=\hsize]{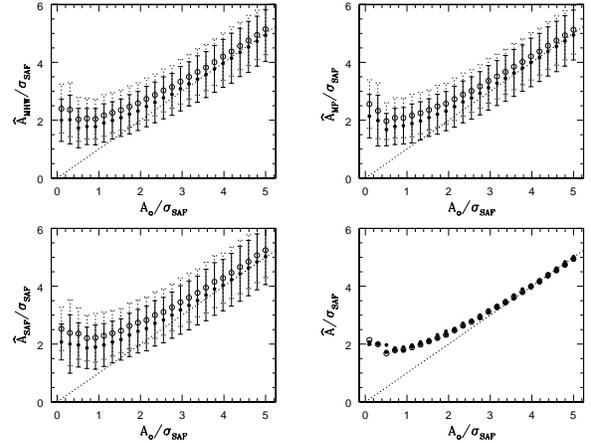}
\caption{Determination of the amplitude in the case of simple
measurement and $\gamma=2.2$. 
The estimates using eqn.~(\ref{eq:estimation_of_the_amplitude})
(filled circles) and 
using the measured intensity (open circles) are shown for the three
filters. The lower right panel 
shows the comparison between the three filters (filled circles for
SAF, open circles for MF, asterisks for MHW). }
\label{fig:se_gamma2p2}
\end{figure}

Figure~\ref{fig:se_gamma2p2} shows the results of the estimation of the
amplitude in the case $\gamma=2.2$. 
There are not qualitative differences with respect to the previous two cases.
Note once more that having the same normalized amplitude $\nu$ 
than in the cases $\gamma = 0, \ 1.4$ does not
mean having the same amplitude at the beginning. As the gain of the
filter grows, we can reach fainter sources.

\section{Conclusions}

 Filtering is a useful technique for the
denoising and enhancing of compact signals in Astronomy and many
other applications.
However, the diversity of different kind of filters
makes difficult to choose which one is the most
convenient in any particular case.
In this paper we have addressed the problem of the choice of
filter for the detection and measurement of point sources 
in
a noisy background.
We focus on the one-dimensional case and assume that the sources
have a Gaussian spatial profile and the noise can be modelled
by an isotropic and homogeneous Gaussian random field.
Then we compare three different filters that have been thoroughly
used  in the study of Cosmic Microwave Background radiation data: 
the Mexican Hat wavelet, the matched filter and the scale-adaptive filter.
Although we have focused in these three filters, the methodology
we describe can be easily applied to any linear filter.

Two main scenarios have been explored: the case in which one
tries to detect a source of known amplitude but uncertain
existence and location in the data (\emph{simple detection}) and
the case in which one tries to detect sources and determine their
a priori unknown amplitude (\emph{simple estimation}). In both
cases, local peak detection is the first step to be undertaken.
In the one-dimensional case,
any local peak is characterised by its intensity (amplitude) $\nu$ and
its curvature $\kappa$. Once a peak is detected, a \emph{decision}
about its validity
must be done. 

We use the Neyman-Pearson criterion to define a region of 
acceptance in the $(\nu,\kappa)$ space that maximizes
the \emph{significance} of the detections. The limits of such region
will depend on the properties of the background (namely, its
power spectrum) and the filter that has been applied to the data.

In the simple measurement scenario, the Neyman-Pearson criterion
is applied by calculating the a posteriori p.d.f. of all the possible values
of the amplitude given the data  $(\nu,\kappa)$. In absence of
any a priori information about the p.d.f. of the amplitudes of the sources
$p(A)$,
we assume uniformity between $0$ and a certain cut $\nu_c$ that can
be easily guessed from the data. If we had a priori information about
$p(A)$, it could be straightforwardly included in the formalism.
Once the source is detected, its amplitude $A$ can be estimated
by maximizing the likelihood $L(\nu, \kappa | \nu_s)$ with
respect to $\nu_s$. Convenient confidence intervals can be given for any
detection using the previous normalized distribution.

We have compared the MHW, the SAF and the MF in the two scenarios
presented above. Both analytical and empirical comparisons have
been performed considering a generic background
characterised by a power spectrum $P(q) \propto q^{-\gamma}$. 
In the analytical comparison 
we have shown how to derive useful formulae 
to predict the number density of local peaks that 
lie inside the Neyman-Pearson region of acceptance 
both in presence and in absence of source. With these quantities
it is possible to compute the ratio between the number of
true detected sources and the number of
spurious detections, that is, a measure of the \emph{reliability}
of each filter in the task of detecting sources.
We find that, regarding this last quantity, there are three
different regions demarcated by the value of the index $\gamma$:
for $0 \leq \gamma < 1$ the MF outperforms the other
two in the reliability sense as well as in total number of 
true detections. For $\gamma=1$ the MF and the SAF coincide.
For $1 <  \gamma \la 1.6$ the SAF gives the best reliability.
The relative difference with the MF is greater than a $10\%$.
For $\gamma=2$ the MF and the MHW coincide.
Finally, for $\gamma \ga 1.6 $ the MHW is the most reliable filter for
the detection, although
the matched filter performs almost equally well. However, the
performance of the MHW can be improved by using it with an optimal
scale (that can be estimated from the data) which gives maximum
amplification of the source. These three regions are present and their limits
are roughly the same both in the simple detection
and in the simple measurement scenarios.
 
We have performed exhaustive simulations in order 
to test the previous ideas.
The cases $\gamma=0$, $1.4$ and $2.2$ have been chosen 
with the aim of
exploring the three regions described above.
The results of the simulations totally agree with the
analytical expected behaviour.  

Regarding the estimation of sources of unknown amplitude,
the three filters perform equally well. The estimation
of the amplitude using the likelihood of the data
is fairly better than the estimation using directly the
measured intensity of the peaks. At low source amplitudes
there is a positive bias due to a selection effect. With these
filters and the likelihood estimator, it is possible to safely 
reach thresholds as low as $2.5 \sigma_{\rm SAF}$ in the filtered
data.  Due to the amplification effect of the filters,
the equivalent thresholds  
in the unfiltered maps can be really low,
specially for high $\gamma$ indexes. As an example,
in the case $\gamma=2.2$ the MF produced in our simulations a
mean gain of $7.5$, this meaning that the 
`safe threshold' of $2.5$ after filtering translates into
a $0.33$ threshold before filtering.

The ideas presented in this paper can be generalized 
to more general filtering schemes and for two-dimensional data
sets. In this case, other quantities such as the ellipticity of
the peaks can be useful to establish decision criteria
similar to the ones presented here. A future work will address
this particular issue.

\section*{Acknowledgements}

The authors thank Patricio Vielva for useful discussions.
RBB thanks the Ministerio de Ciencia y Tecnolog\'\i a and the
Universidad de Cantabria for a Ram\'on y Cajal contract.
DH acknowledges support from a UC postdoctoral fellowship during
the months of March to October 2002 and from the
European Community's Human Potential Programme
under contract HPRN-CT-2000-00124, CMBNET, since 
November $1^{st}$ 2002. We acknowledge partial support from 
the Spanish MCYT projects ESP2001-4542-PE and ESP2002-04141-C03-01
and from the EU Research Training Network `Cosmic Microwave Background
in Europe for Theory and Data Analysis'.

\end{document}